\documentclass{aa}  
\usepackage{txfonts}

\usepackage{longtable}

\usepackage{makecell}
\usepackage{graphicx}
\usepackage{natbib}
\usepackage[utf8]{inputenc}
\usepackage{natbib}
\usepackage[mathscr]{euscript}
\usepackage{lscape}

\usepackage{booktabs,caption,fixltx2e}
\usepackage[flushleft]{threeparttable}
\usepackage[dvipsnames]{xcolor}
\definecolor{myblue}{RGB}{63,200,244}
\definecolor{ocre}{RGB}{52,177,201}

\begin{document}

   \title{Neutron-capture elements in dwarf galaxies~I:\\Chemical clocks \& the short timescale of the $r$-process\thanks{Based on VLT/FLAMES observations collected at the European Organisation for Astronomical Research (ESO) in the Southern Hemisphere under programmes 71.B-0641, 171.B-0588 and 092.B- 0194(A).}\thanks{Tables B.1 and B.2 are available in electronic form
at the CDS.
}}

   \author{\'{A}.~Sk\'{u}lad\'{o}ttir
   		\inst{1}
        \and
        C.~J.~Hansen\inst{1}
       	\and
        S.~Salvadori \inst{2}   
        \and	   	
       	A.~Choplin\inst{3}
       	}

   \institute{
   			Max-Planck-Institut f$\ddot{\text{u}}$r Astronomie, K$\ddot{\text{o}}$nigstuhl 17, D-69117 Heidelberg, Germany\\
   			\email{skuladottir@mpia.de}
   			 \and
   			 Dipartimento di Fisica e Astronomia, Universit\'{a} degli Studi di Firenze, Via G. Sansone 1, I-50019 Sesto Fiorentino, Italy
   			 \and
   			 Department of Physics, Faculty of Science and Engineering, Konan University, 8-9-1 Okamoto, Kobe, Hyogo 658-8501,    	Japan
                }


\abstract{The heavy elements ($Z>30$) are created in neutron ($n$)-capture processes that are predicted to happen at vastly different nucleosynthetic sites. To study these processes in an environment different from the Milky Way, we targeted the $n$-capture elements in red giant branch stars in the Sculptor dwarf spheroidal galaxy. Using ESO VLT/FLAMES spectra, we measured the chemical abundances of Y, Ba, La, Nd, and Eu in 98 stars covering the metalliticy range $-2.4<\text{[Fe/H]}<-0.9$. This is the first paper in a series about the $n$-capture elements in dwarf galaxies, and here we focus on the relative and absolute timescales of the slow ($s$)- and rapid ($r$)-processes in Sculptor. From the abundances of the $s$-process element Ba and the $r$-process element Eu, it is clear that the $r$-process enrichment occurred throughout the entire chemical evolution history of Sculptor. Furthermore, there is no evidence for the $r$-process to be significantly delayed in time relative to core-collapse supernovae. Neutron star mergers are therefore unlikely the dominant (or only) nucleosynthetic site of the $r$-process. However, the products of the $s$-process only become apparent at $\text{[Fe/H]}\approx-2$ in Sculptor, and the $s$-process becomes the dominant source of Ba at $\text{[Fe/H]}\gtrsim-2$. We tested the use of [Y/Mg] and [Ba/Mg] as chemical clocks in Sculptor. Similarly to what is observed in the Milky Way, [Y/Mg] and [Ba/Mg] increase towards younger ages. However, there is an offset in the trends, where the abundance ratios of [Y/Mg] in Sculptor are significantly lower than those of the Milky Way at any given age. This is most likely caused by metallicity dependence of yields from the $s$-process, as well as by a different relative contribution of the $s$-process to core-collapse supernovae in these galaxies. Comparisons of our results with data of the Milky Way and the Fornax dwarf spheroidal galaxy furthermore show that these chemical clocks depend on both metallicity and environment.
}               
               
                \maketitle

%
\section{Introduction}

Chemical elements heavier than zinc are created in processes where a seed nucleus undergoes neutron ($n$)-capture and consequent $\beta$-decays to form an isotope with a higher atomic mass \citep{Burbidge57,Sneden08,Nomoto13,Frebel18}. This nuclosynthesis occurs in the slow ($s$), intermediate ($i$), and rapid ($r$) processes, depending on $n$-density. In addition, the lighter-element primary process (LEPP), also called \textit{`weak'} or \textit{`limited'} $r$-process, has been invoked to explain the abundances of the lighter $n$-capture elements (e.g., Sr, Y, and Zr) at low metallicities in the Milky Way (e.g., \citealt{Travaglio04,Francois07}). Most heavy elements are formed in several or all of these processes, but in different proportions and isotope ratios depending on the underlying physical conditions. From an observational point of view, Ba is often viewed as the canonical tracer of the $s$-process, because it is relatively straightforward to measure, and $\sim$85\% of the solar abundance comes from this nucleosynthetic channel. Similarly, Eu is the most commonly used tracer of the $r$-process, which forms $\sim$94\% of the solar abundance \citep{Bisterzo14}.

\subsection{The $s$-process and chemical clocks}

The $s$-process occurs in the asymptotic giant branch (AGB) phase of low- to intermediate-mass stars, $\lesssim10$~M$_\odot$, during the last $\sim$1\% of their lifetimes \citep{Herwig05,Karakas14,Frebel18}. The heavy elements from the $s$-process are released into the environment through stellar winds, with time delays depending on the lifetimes (and thus masses) of the stars. This time delay causes the abundances of $s$-process elements to increase with time relative to the products of core-collapse supernovae (ccSN). In particular, [Y/Mg] has been shown to have a very tight correlation with age in solar twins \citep{daSilva12,Nissen15,Nissen16,Nissen17,TucciMaia16,Spina18}. Furthermore, the same trends of [Y/Mg] with age were observed in solar metallicity giant stars in four open clusters \citep{Slumstrup17}. 

Accurate stellar ages are notoriously challenging to measure (e.g.,~\citealt{Soderblom10}). Therefore it would be very useful to find abundance ratios that could be used as a proxy for age (e.g.,~\citealt{Nissen18,Silva18}). Although [Y/Mg] is very promising as such a \textit{`chemical clock'}, observations indicate that the correlation with age might depend on metallicity \citep{Feltzing17} and/or environment, because some differences are seen between the thick and thin disks \citep{DelgadoMena19,Titarenko19}. This raises the question of how universal these chemical clocks are, especially because they have currently not been studied in other galaxies.

\subsection{The $r$-process}

The astrophysical site(s) of the $r$-process remain uncertain. High-energy, neutron-rich environments are required, so that the proposed sites are typically connected with violent events, such as SN explosions or the mergers of compact objects (e.g.,~\citealt{Sneden08,Nomoto13}). The empirically determined chemical abundance pattern of the $r$-process is very robust for a wide range of elements, $55<Z<72$; $r$-process rich low-metallicity stars are consistent with the solar scaled $r$-abundance (e.g.,~\citealt{Hill02,Hill17,Sneden03,Sneden08}). Furthermore, the large scatter of [Eu/Fe] at low metallicities, $\text{[Fe/H]}\lesssim-2.5$, in the Milky Way halo (e.g.,~\citealt{Francois07}) suggests a rare and prolific source.

The detection with the Laser Interferometer Gravitational-Wave Observer (LIGO) of a neutron star merger (NSM), GW170817, and the following observations of ultraviolet, optical and infrared emission are consistent with $r$-process nucleosynthesis \citep{Abbott17b,Abbott17a,Chornock17,Cowperthwaite17,Drout17,Pian17,Tanaka17,Villar17}. Furthermore, the estimated rate of NSM extrapolated from this single event, has been shown to be sufficient to account for the production of all $r$-process material in the Milky Way (e.g., \citealt{Chornock17,Cowperthwaite17,Hotokezaka18,Rosswog18}). However, the uncertainties of the rate, nucleosynthetic yields and delay times of NSMs remain large, and the existence of NSM does of course not automatically exclude other sources (\citealt{Cote19}).

The host galaxy of the NSM GW170817 has been revealed to be an early-type galaxy \citep{Abbott17c,Coulter17} with star formation rate $\lesssim10^{-2}$~M$_\odot$\,yr$^{-1}$ and a predominantly old stellar population \citep{Blanchard17,Levan17,Pan17}. Based on the analysis of \citet{Pan17}, the delay time of this NSM is estimated to be $\gtrsim3$~Gyr, while \citet{Blanchard17} estimated a 90\% probability of the time delay to be between 6.8 and 13.6~Gyr. Overall the properties of the host galaxy are consistent with those of short gamma-ray bursts (GRB; \citealt{IM17}). The progenitors of short GRBs are believed to be compact object binary mergers, and around one third of those are found in early-type galaxies \citep{Berger14,Fong17}. Thus, a significant fraction of compact object binaries must take a long time to merge (see also the more detailed discussion in \citealt{Cote19}). 


Given the available observations, NSM are a very promising site for the $r$-process. However, the abundance pattern of stars in the Milky Way has proven difficult to model when they are assumed to be the dominant (or only) source. The detection of $r$-process elements in stars at the lowest metallicities (e.g.,~\citealt{Francois07}) would require that a fraction of NSM occurs very soon after star formation starts. Some chemical evolution models require minimum timescales on the order of $\sim$1-10~Myr \citep{Matteucci14,Cescutti15}, while other have been able to reproduce the abundances of [Eu/Fe] at $\text{[Fe/H]}<-2$ with minimum timescales of $\sim$100~Myr \citep{Shen15,Hirai15,Ishimaru15}.

Furthermore, when physically motivated time-delay distributions are used (instead of a fixed delay time), the decreasing trend of [Eu/Fe] with $\text{[Fe/H]}>-1$ is challenging to reproduce assuming NSM as the only $r$-process source \citep{Shen15,Voort15,Komiya16,Cote17,Cote19,Hotokezaka18,Simonetti19}, although some simulations have been successful \citep{Naiman18}. Several additional mechanisms have been invoked to explain this discrepancy, such as natal kicks \citep{Tauris17}, metallicity dependence \citep{Simonetti19} or details in the physics of the interstellar medium (ISM; \citealt{Schonrich19}). Overall there are still significant discrepancies between population synthesis models of NSM and Galactic chemical evolution models, under the assumption that NSM are the only source of the $r$-process \citep{Cote17}.

Another possibility is that an additional $r$-process source is associated with a rare type(s) of ccSN. \citet{Winteler12} proposed a magnetohydrodynamic (MHD) driven SN with jets and strong magnetic fields as a possible $r$-process source. However the 3D simulations of \citet{Mosta18} suggest that under realistic assumptions about the magnetic field, MHD SN might not be a major source of $r$-process elements. 
Recent simulations predict that \textit{`collapsars'}, that is, the collapse of massive stars, can produce all of the $r$-process material in the Universe \citep{Siegel19a,Siegel19b}. The accretion disks around such events yield a very efficient $r$-process machine, and the authors suggest that the $r$-process arising from NSM GW170817 originated in a similar accretion disk around the black hole that formed in the merger. Furthermore, the $r$-process could occur in a limited mass range, 8-10~M$_\odot$, of low-mass SN \citep{Ishimaru04,Ishimaru05,Wanajo09}. All these proposed scenarios remain uncertain and have very limited observational constraints.

       \begin{figure}
   \centering
   \includegraphics[width=\hsize-0cm]{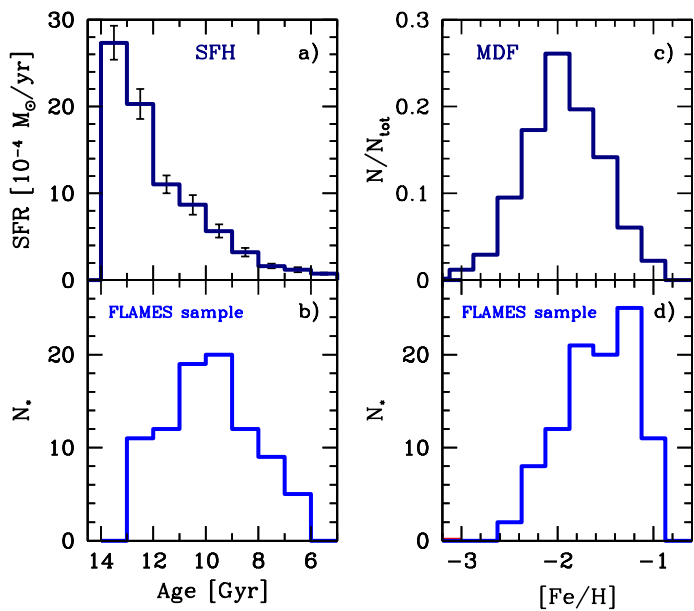} 
      \caption{a) SFH of Sculptor \citep{deBoer12} b) The ages of our target stars where available (88 stars) from \citet{deBoer12} c) MDF of Sculptor \citep{Battaglia08a}; d) MDF for our target stars.
      }
         \label{fig:sfhmdf}
   \end{figure}

\subsection{The Sculptor dwarf spheroidal galaxy}

The Sculptor dwarf spheroidal (dSph) galaxy is a well-studied satellite galaxy of the Milky Way at a distance of $86\pm5$~kpc \citep{Pietrzynski08}. This galaxy is dominated by an old stellar population $>10$~Gyr, see~Fig.~\ref{fig:sfhmdf}, with a total stellar mass $\sim$10$^6$~M$_\odot$ \citep{deBoer12}. High-resolution (HR) spectroscopic analysis of some heavy element abundances in Sculptor has been performed by different authors for $\leq5$ stars each \citep{Shetrone03,Geisler05,Tafelmeyer10,Frebel10,Kirby12,Skuladottir15a,Jablonka15,Simon15}, and for 99 stars in \citet{Hill18}. Recently, Ba in this galaxy has also been studied by \citet{Duggan18}, and \citet{Starkenburg13} included Sr and Ba measurements and upper limits for seven stars at low metallicity, $\text{[Fe/H]}<-3$, using spectra of intermediate resolution. For a general and more complete overview of the previous studies of Sculptor, see \citet{Hill18} and references therein.

In this paper, we present new heavy elemental abundances for 98 stars in the Sculptor dSph galaxy. The abundances of various elements of our target stars have been studied in many previous works \citep{Tolstoy09,North12,Skuladottir15b,Skuladottir17,Skuladottir18,Hill18}. Here we present an independent analysis of Mg and the heavy elements Y, Ba, La, Nd, and Ba. This first paper in a series about the $n$-capture elements in dwarf galaxies focuses on the relative timescales of the $s$- and $r$-process in Sculptor. A second paper, \textit{Neutron-capture elements in dwarf galaxies~II: Challenges for the s- and i-processes at low metallicity} (hereafter Paper~2; Sk\'{u}lad\'{o}ttir et al. in prep.), will discuss the implications of these new data for the $s$-process and LEPP in Sculptor. The third paper in the series, \textit{Neutron-capture elements in dwarf galaxies III, A homogenized analysis of 13 dwarf spheroidal and ultra-faint galaxies}  (hereafter Paper~3; Reichert et al. in prep.), will present a homogeneous abundance analysis of $n$-capture elements in dwarf galaxies, using archival data.

The well-understood star formation history (SFH) of Sculptor allows the determination of precise stellar ages \citep{deBoer12}. Thus we can study the $n$-capture processes in a system that has a very different chemical enrichment history from the Milky Way. Using Sculptor as a laboratory allows us to address some of the most pressing questions of the $s$- and $r$-processes, such as their time-delay distribution and the effects of environment on the enrichment processes.

\section{Abundance analysis}

\subsection{Target sample}

The stellar sample consists of 98 red giant branch (RGB) stars, which were previously identified as members of the Sculptor dSph by \citet{Hill18}. These observations were taken with ESO VLT FLAMES/GIRAFFE (88 stars) and FLAMES/UVES (10 stars). In addition to the spectra presented in \citet{Hill18}, we also include the FLAMES/GIRAFFE spectra taken with the HR7A setting from \citet{Skuladottir17}, for 84 stars in common between the two (GIRAFFE) surveys. Four stars in the sample therefore do not have HR7A spectra. The wavelength ranges, resolution and observing times of the different settings are listed in Table~\ref{tab:FLAMES}. For further details about the observations, Sculptor membership and the data reduction, see \citet{Skuladottir17} and \citet{Hill18}.

The metallicity distribution function (MDF) of stars in the Sculptor dSph peaks around $\text{[Fe/H]}\approx-2$. However, our stellar sample is selected in the center of the galaxy, and is thus biased toward higher metallicities, see Fig.~\ref{fig:sfhmdf}. The low-metallicity tail of the galaxy, $\text{[Fe/H]}<-2.5$, is not included in our sample, but has been studied elsewhere \citep{Tafelmeyer10,Frebel10,Starkenburg13,Jablonka15,Simon15,Chiti18}. Similarly, our target sample is biased toward younger stars because they mainly reside in the center of the galaxy \citep{Tolstoy04,deBoer12}. While the majority of stars in Sculptor are >10~Gyr old, the age distribution of our sample peeks around $\sim$10~Gyr and the (rare) youngest population is well represented.

\begin{table}
\caption{Wavelength range, resolution, and observing time of the GIRAFFE and UVES settings used here.
}
\label{tab:FLAMES}
\centering
\small
\begin{tabular}{l c c c c}
\hline\hline
Setting & $\lambda_\textit{min}$ & $\lambda_\text{max}$ & Resolution & Obs. time \\
& [$\AA$] & [$\AA$] & &\\
\hline
HR7A    & 4700 & 4970 & 19\,500 & 6hr\\ 
HR10	 & 5340 & 5620 & 19\,800 & 4hr30min\\
HR13	 & 6120 & 6400 & 22\,500 & 4hr20min\\
HR14A	 & 6390 & 6620 & 28\,800 & 7hr\\
HR15	 & 6610 & 6960 & 19\,300 & 2hr\\
UVES     & 4800 & 6800 & 47\,000 & 7 and 11hr\\
\hline\hline
\end{tabular}
\end{table}

\subsection{Stellar parameters and abundance analysis}

The stellar parameters, temperature $T_\textsl{eff}$, surface gravity $\log{g}$, microturbulence velocity $v_t$, and [Fe/H], are adopted from \citet{Hill18} and are listed in Table~\ref{tab:abundances}. These stellar parameters have also previously been used by \citet{North12} and \citet{Skuladottir15b,Skuladottir17}. For the abundance analysis, the stellar atmosphere models are adopted from MARCS\footnote{marcs.astro.uu.se} \citep{Gustafsson08} for stars with standard composition, 1D, and assuming local thermodynamic equilibrium (LTE), interpolated to match the stellar parameters for the target stars. The abundance analysis was made with the spectral synthesis code TURBOSPEC\footnote{ascl.net/1205.004} \citep{AlvarezPlez98,Plez12}. The continuum evaluation was carried out with a synthetic spectra analysis, independently of that of \citet{Hill18}. Atomic parameters were adopted from the VALD\footnote{http://vald.astro.uu.se} database (\citealt{Kupka99} and references therein). All lines we used for abundance measurements are listed in Table~\ref{tab:linelist}. Measurements were made by including all atomic data for the wavelength range in question, thus including blends of other elements.

The errors were evaluated as is described in \citet{Skuladottir17}. When four or more lines were measured for a given element in a star, the final abundance was defined as the average, with the error of the mean. On the other hand, when only three or fewer lines were available, the abundance was defined as the weighted average, and the errors were weighted accordingly (for details, see \citealt{Skuladottir15a}). The systematic errors from the stellar parameters were not included, but in almost all cases, the error on the abundance ratio [X/Y] of two metals is dominated by the line measurement error. For a representative error on [X/Y] due to uncertainties in stellar parameters, see \citet{Hill18}.

For consistency with previous work on this stellar sample, we adopted the solar abundances from \citet{GrevesseSauval98}, A(Fe)$_\odot=7.50$, A(Mg)$_\odot=7.58$, A(Y)$_\odot=2.24$, A(Ba)$_\odot=2.13$, A(La)$_\odot=1.17$, A(Nd)$_\odot=1.50$, and A(Eu)$_\odot=0.51$. Literature data are adjusted to this scale.

\subsection{Individual elements} \label{sec:ind}

To minimize systematic errors, the elemental abundances for Mg, Y, Ba, La, Nd, and Eu were all evaluated with the same method, and are presented in Table~\ref{tab:abundances}. Owning to the low signal-to-noise ratio (S/N) at $\lesssim4750~\AA$ in the HR7A GIRAFFE spectrum (see \citealt{Skuladottir17}), only lines at $\gtrsim4800~\AA$ were used. The line list is given in Table~\ref{tab:linelist}. When available, the previously determined elemental abundances by \cite{Hill18} agree with those presented here; for a more detailed discussion see Appendix~A.

The magnesium abundances were derived as a representative $\alpha$-element, using one \ion{Mg}{I} line, at $5528.4~\AA$, in the GIRAFFE sample (HR13). For the UVES sample two additional lines were also used, at $5183.6$~and $5711.0~\AA$. The Mg abundance could be reliably measured in all stars with the exception of ET0299, where combined with relatively low metallicity and low S/N, a spike at the center of the line in its GIRAFFE spectrum hindered reliable determination. 

The light $n$-capture element yttrium was measured using three \ion{Y}{II} lines at $4850-4900~\AA$. Therefore it could not be measured in the four stars with no HR7A spectrum (ET0009, ET0013, ET0035 and ET0039), see Table~\ref{tab:abundances}, or in a spectrum in this region of extremely poor quality (ET0342). With the higher resolution and longer wavelength coverage of UVES, up to eight lines could be used for the abundance determination of \ion{Y}{II}, see Table~\ref{tab:linelist}. 

Barium was measured using three \ion{Ba}{II} lines, at $5853.7$, $6141.7$, and $6496.9~\AA$, where the bluest of the three was only accessible in the UVES spectra. Ba could be reliably measured in all stars, but because of defects in the spectra, only one line was used in the stars ET0198, ET0202, and ET0376.

In total eight \ion{La}{II} lines were used for the GIRAFFE spectra, three in the bluest setting (HR7A) and five in the red (HR14A and HR15). In the UVES spectra eight additional lines were also used. Abundances for La were measured in the majority of the sample, 72 GIRAFFE and 8 UVES stars. In almost all cases where La was undetected, the metallicity of the stars was very low $\text{[Fe/H]}\lesssim-2$ and/or the HR7A region was missing. 

We used about 20 lines to measure neodymium in the GIRAFFE spectra, and about 40 in the UVES spectra. The Nd abundances were measured for 71 GIRAFFE stars, and 10 UVES spectra. Out of the 13 stars where we were unable to obtain Nd abundances, 12 had $\text{[Fe/H]}\lesssim-2$.

Only one \ion{Eu}{II} line was accessible in the spectra, at $6645.1~\AA$. This is a weak line, and was detected in about half of our sample, 44 GIRAFFE and 7 UVES stars. It was typically detected in stars with average or above average S/N and $\text{[Fe/H]}>-2.2$.

In addition to the elements mentioned above (Y, Ba, La, Nd, and Eu), the spectra were examined for visible lines of other heavy elements (e.g., Ce). The quality of the spectra was not sufficient for a reliable abundance determination of these elements.

          \begin{figure*}
   \centering
   \includegraphics[width=\hsize-2cm]{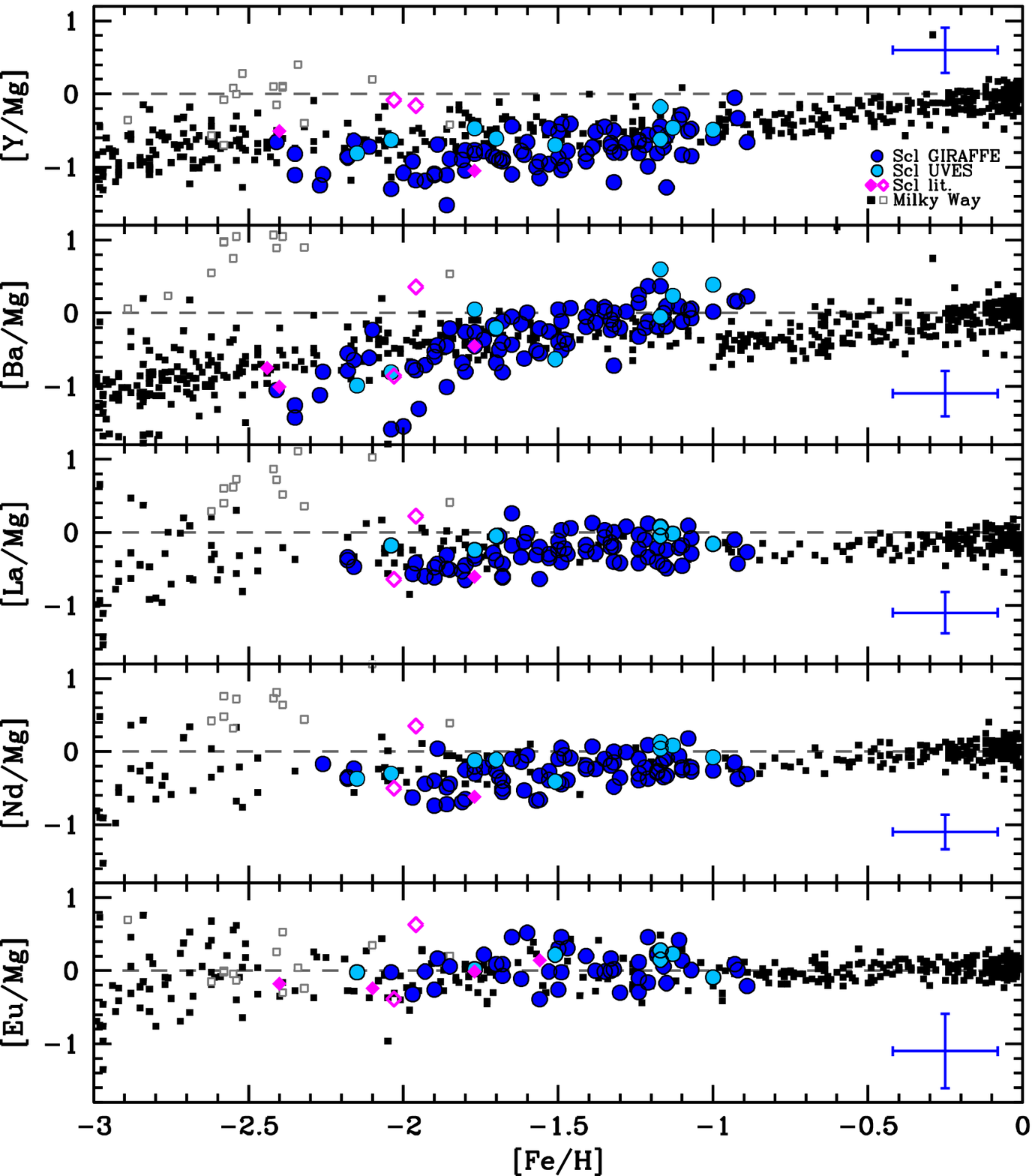}
      \caption{Ratios of the $n$-capture elements to Mg. Target stars are shown with blue (GIRAFFE) and light blue (UVES) circles. The representative error bar for the Sculptor data is shown in blue. Previous measurements in Sculptor from HR spectra are shown with magenta diamonds \citep{Shetrone03,Geisler05,Kirby12,Skuladottir15a,Jablonka15}. Open diamonds are stars with peculiar abundances of the $n$-capture elements. Milky Way stars are shown with black squares. Open squares refer to Milky Way stars with high $\text{[Ba/Mg]}>0$ and confirmed high $\text{[C/Fe]}>0.7$. \textit{Milky Way references}: \citealt{Reddy03,Reddy06,Venn04,Francois07,Mishenina13,Roederer14}. The SAGA database \citep{Suda08} was used to gather this compilation.
      }
         \label{fig:ncapmg}
   \end{figure*}

          \begin{figure*}
   \centering
   \includegraphics[width=\hsize-2cm]{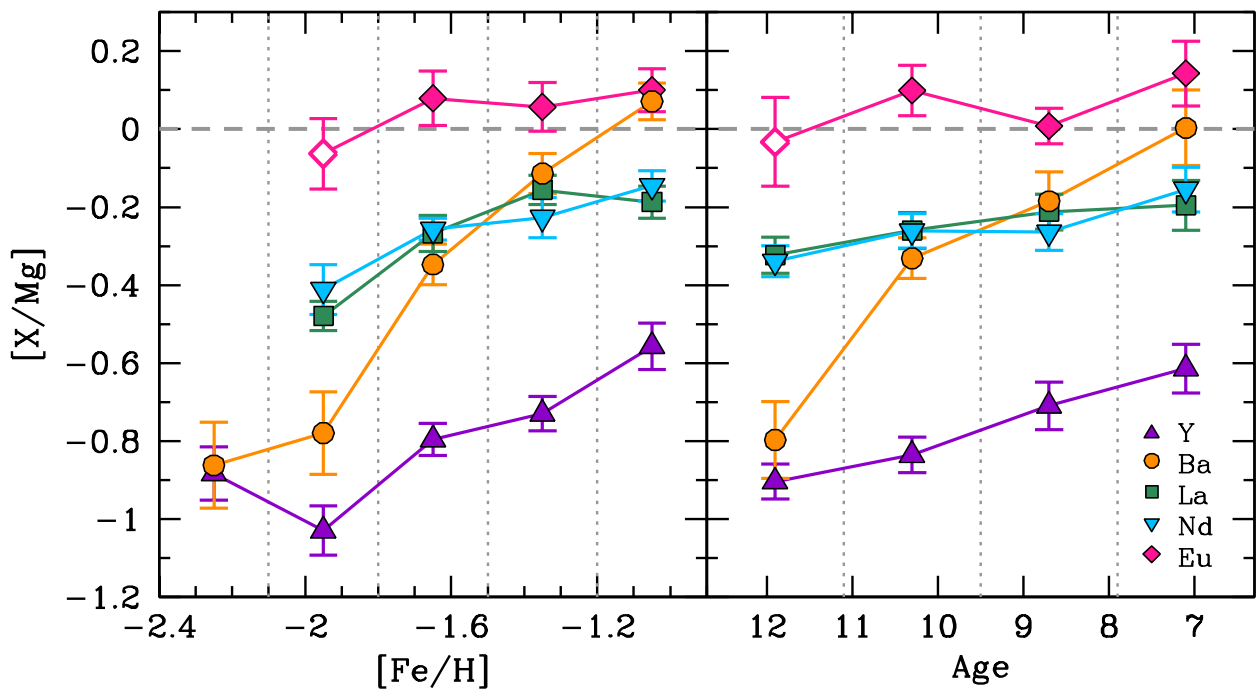}
      \caption{Average [X/Mg] for our sample in five [Fe/H] bins (left panel) and four age bins (right panel): Y (purple triangles), Ba (orange circles), La (green squares), Nd (light blue downward pointing triangles), and Eu (pink diamonds). Open symbols note bins with $<7$~stars, while all filled points contain $\geq10$~stars. Dashed vertical lines show the sizes of the bins, and the y-error bar is the error of the mean. 
      }
         \label{fig:mgplot}
   \end{figure*}

                   \begin{figure*}
   \centering
   \includegraphics[width=\hsize-6cm]{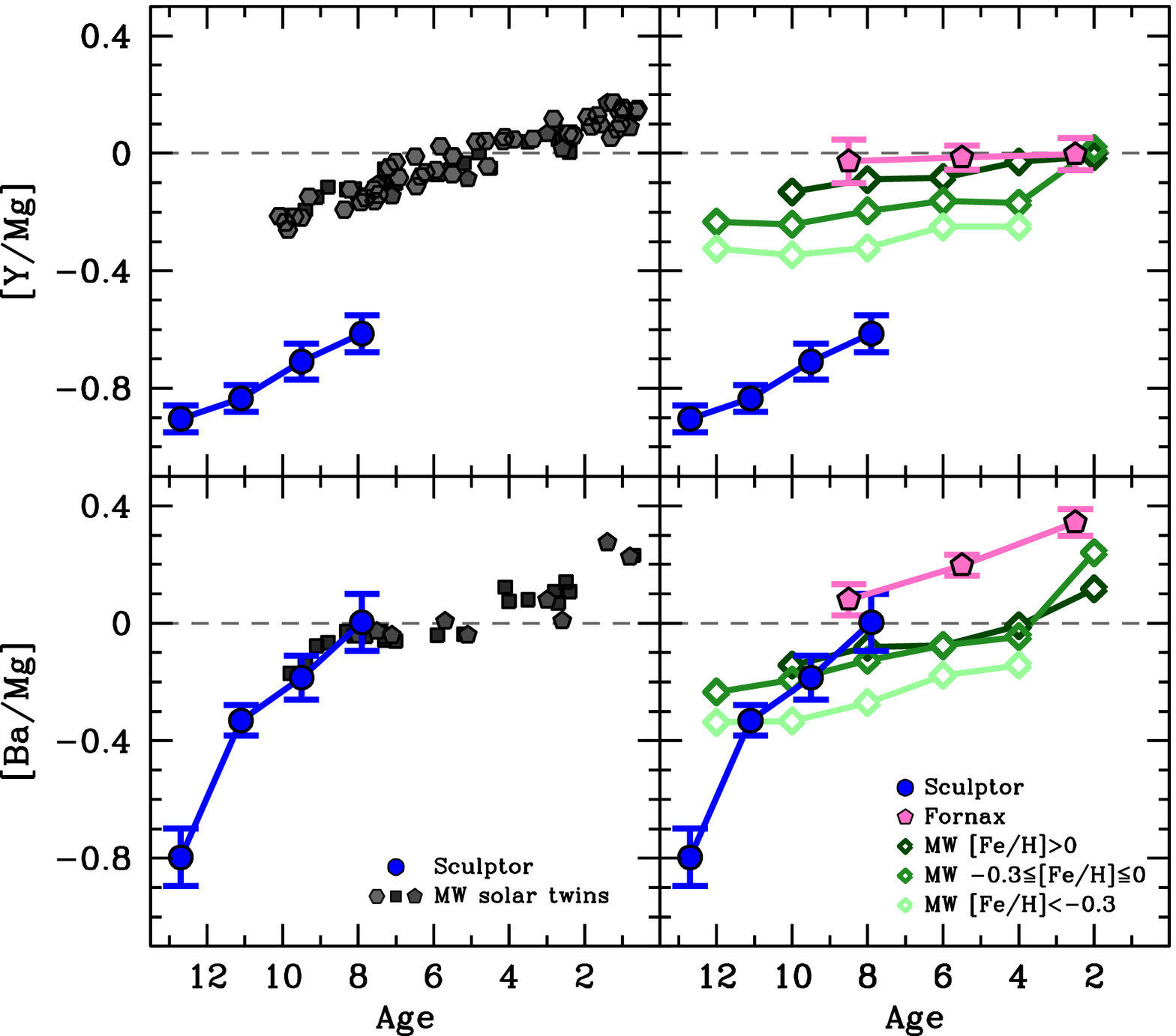}
      \caption{ [Y/Mg] and [Ba/Mg] as a function of age. Sculptor binned data are blue circles. \textit{Left panels:} Comparison with individual Milky Way solar twin stars (gray symbols). \textit{Right panels:} Comparison with binned data from the Fornax dSph (pink pentagons) and the Milky Way (open diamonds) for three metallicity bins:  i)~$\text{[Fe/H]}>0$ (dark green), ii)~$-0.3\leq\text{[Fe/H]}\leq 0$ (green), and iii)~$\text{[Fe/H]}<-0.3$ (light green). In the case of the Milky Way, the error of mean is smaller or comparable to the symbols shown. For the dSphs, the error of the the mean is shown with a y-error bar.
\textit{MW references:} \citealt{Nissen15,Nissen16,Nissen17} (squares); \citealt{TucciMaia16} (hexagons); \citealt{Spina18} (pentagons); \citealt{Bensby14} (open diamonds). \textit{Fornax references:} \citealt{deBoer12f} (ages); \citealt{Letarte10,Letarte18} (Mg, Y); \citealt{Andrievsky17} (Ba).       
      }
         \label{fig:age}
   \end{figure*}

\section{Chemical clocks}

\subsection{Timescales of the $r$- and $s$-processes}

The interpretation of the abundance ratios [X/Fe] with [Fe/H] and/or time are complicated by the Fe that is produced by SNe type~Ia, especially because their influence becomes apparent at different metallicities in Sculptor and the Milky Way \citep{Tolstoy09,Skuladottir15b,Hill18}. Because SNe type~Ia do not affect the abundances of the $n$-capture elements, it is preferable to avoid the unnecessary complication of using Fe as reference element in the abundance ratios.\footnote{For the trends of heavy elemental abundances [X/Fe] with [Fe/H] in Sculptor see \citet{Hill18}, and Paper~2.} In Fig.~\ref{fig:ncapmg} we therefore plot [X/Mg] versus [Fe/H] because Mg is a very good tracer of ccSN. In this particular sample in the center of Sculptor, [Fe/H] is a reasonable proxy for age (see \citealt{Hill18}), and in the following we therefore assume that [Fe/H] is a good indicator of the evolutionary state of the galaxy when the stars were formed.

When we focus on the general population in the Milky Way (black squares in Fig.~\ref{fig:ncapmg}) it is evident that both [Y/Mg] and [Ba/Mg] increase with [Fe/H] and the same is true in Sculptor. This is consistent with a delayed contribution of Y and Ba relative to ccSNe, as expected from the $s$-process. The trends in the Sculptor dSph and the Milky Way are qualitatively similar. However, the abundances of [Y/Mg] in Sculptor are on average lower than those of the Milky Way at the same metallicities, and [Ba/Mg] has a different slope in these two galaxies. As the star formation histories are vastly different, and thus also the amount of AGB contribution at a given metallicity, these differences are expected. 

The elements La and Nd are predicted to have less contribution from the $s$-process than Ba (e.g., \citealt{Bisterzo14}). In agreement with this generally accepted premise, [La/Mg] and [Nd/Mg] only mildly increase with [Fe/H] both in Sculptor and the Milky Way, see Fig.~\ref{fig:ncapmg}. There is a slight indication that these ratios are not increasing at $\text{[Fe/H]}\lesssim-2$ in Sculptor. However, these elements become very challenging to measure at the lowest metallicities with our spectra, so the available measurements might be biased towards higher values.

The $r$-process element Eu has a flat trend of [Eu/Mg] with [Fe/H] around the solar value in Sculptor, see Fig.~\ref{fig:ncapmg}, with a measured slope of $0.03\pm0.10$ per dex in [Fe/H]. Similarly, the Milky Way has an average of $\text{[Eu/Mg]}\approx0$ at all [Fe/H] with increasing scatter toward the lowest metallicities. If there were a significant delay in the enrichment of Eu compared to Mg (or vice versa), increasing (or decreasing) trends would be expected in their ratios with metallicity, as is seen in case of the $s$-process (e.g., [Y/Mg] and [Ba/Mg]; Fig.~\ref{fig:ncapmg}) and SN type~Ia (e.g., [Cr/Mg] and [Fe/Mg]; \citealt{Hill18}). The absence of such trends suggests that the timescales of Eu and Mg production are similar, that is, that \textit{the dominant source of Eu enriches the ISM with timescales comparable to those of massive stars.} Furthermore, the similarities between the Sculptor dSph and the Milky Way, suggest that the [Eu/Mg] is not significantly affected by differences in the SFHs and the environment of these two vastly different galaxies. For further discussion of the main $r$-process source in Sculptor see Section~\ref{sec:global}.

Metal-poor Milky Way stars that have high $\text{[Ba/Mg]}>0$ and are confirmed to be carbon-enhanced metal-poor (CEMP) stars are marked with open squares in Fig.~\ref{fig:ncapmg}. Most (if not all) of these stars are expected to have experienced mass-transfer from a binary companion \citep{Lucatello05,Starkenburg14,HansenT16s}. These CEMP stars are thus not good tracers of the general enrichment of the ISM and we therefore do not discuss them further in this context. More details about the comparison of these stars with the Sculptor data can be found in Paper~2. 

The Sculptor abundance ratios from Fig.~\ref{fig:ncapmg} are shown with binned data in Fig.~\ref{fig:mgplot}, both as a function of [Fe/H] (left) and stellar ages (right). Ages are available for 88 stars in our sample (\citealt{deBoer12}), with an average error of $\langle\delta_\textsl{age}\rangle=1.8$~Gyr. In spite of the reduced sample size and significant errors, the average trends of [X/Mg] with age are comparable with those as a function of [Fe/H], showing that, on average, [Fe/H] is a good indicator for the time when the star is formed, in this particular sample. 

As we discussed in relation to Fig.~\ref{fig:ncapmg}, [Ba/Mg] shows the strongest increase with [Fe/H], followed by [Y/Mg] which also has a very clear trend, see Fig.~\ref{fig:mgplot}. The increases of [La/Mg] and [Nd/Mg] with [Fe/H] are significally less steep and [Eu/Mg] is consistent with a flat trend. Because [Y,Ba/Fe] are consistent with a flat trend at $\text{[Fe/H]}<-2$, we can reasonably infer that the contribution from the $s$-process only becomes significant at $\text{[Fe/H]}\approx-2$ (corresponding to $\sim$11.5~Gyr ago). As both Y and Ba were measured in all stars with the necessary wavelength coverage (see Sec.~\ref{sec:ind} and Table~\ref{tab:abundances}), nondetections do not bias this result toward higher values at lower metallicities.

\subsection{Chemical clocks in different environments}   

The abundance ratios [Y/Mg] and [Ba/Mg] have a particularly clear correlation with stellar age in Sculptor, as is shown in Fig.~\ref{fig:mgplot}. A similar result has previously been observed in the Milky Way, where solar twins show an exceptionally good correlation between [Y/Mg] and age, as well as [Ba/Mg] and age \citep{daSilva12,Nissen15,Nissen16,Nissen17,Nissen18,TucciMaia16,Spina18}. This has lead to the discussion of using these abundance ratios as \textit{`chemical clocks'} because accurate stellar ages are notoriously challenging to measure (e.g., \citealt{Soderblom10,Chaplin13}). Empirical relations have been found between these (and other) abundance ratios extending beyond solar twins (e.g., \citealt{DelgadoMena19}). However, by using data of F and G dwarf stars in the solar neighborhood from \citet{Bensby14}, \citet{Feltzing17} have shown that the trend of [Y/Mg] is dependent on metallicity. 

For the first time, we now compare these correlations in the Milky Way with results from another galaxy. The left panels of Fig.~\ref{fig:age} show the abundance ratios of [Y/Mg] and [Ba/Mg] in Sculptor and in Milky Way solar twins. It is clear immediately that although [Y/Mg] increases with age in both galaxies, the trend in Sculptor is significantly offset compared to that of the Milky Way solar twins: it is at much lower values. This is the result of the metallicity dependence of AGB yields, which at lower metallicities have much lower [Y/Ba] than at solar [Fe/H] (e.g., \citealt{Karakas14}). On the other hand, the abundance ratios of [Ba/Mg] with age are in good agreement between Sculptor and the Milky Way solar twins, where the ages overlap. 

The picture becomes more complicated, however, when we compare the Sculptor abundances with those of the Milky Way from \citet{Bensby14} in the right panels of Fig.~\ref{fig:age}. This sample has a broad range in metallicity, $-2.4\lesssim~\text{[Fe/H]}\lesssim~+0.4$, therefore we divided it into three metallicity bins: i)~high $\text{[Fe/H]}>0$ ii)~medium $-0.3\leq\text{[Fe/H]}\leq 0$, and iii)~low $\text{[Fe/H]}~<~-0.3$. The average abundances of [Y,Ba/Mg] in the Milky Way become lower with decreasing metallicity, escpecially in the case of [Y/Mg] and $\text{age}\geq4$~Gyr. However, even in the lowest metallicity data from \citet{Bensby14}, the abundances of [Y/Mg] are higher than those observed in Sculptor. This is perhaps unsurprising, as there is still a significant metallicity difference between these samples; $\langle\text{[Fe/H]}_\textsl{MW-low}\rangle~=~-0.65\pm0.35$ in the lowest metallicity bin of the \citet{Bensby14} data, while in the sample of [Y/Mg] with age in Sculptor, $\langle\text{[Fe/H]}_\textsl{Scl}\rangle=-1.58\pm0.37$. Here errors represent $1\sigma$ of the scatter.

The same tendency of decreasing values with metallicity is also apparent in [Ba/Mg] in the Milky Way. However, in this case, the Sculptor abundances agree quite well with the higher metallicity bins ($\text{[Fe/H]}\geq-0.3$), while they are inconsistent with the values at the lowest metallicity in the Milky Way sample ($\text{[Fe/H]}<-0.3$) for the youngest Sculptor stars. This indicates that the offset of [Y/Mg] and [Ba/Mg] in Sculptor with the Milky Way data is more complicated than a simple metallicity dependence. 

This is confirmed when the the results of Sculptor and the Milky Way are compared to the Fornax dSph, see Fig.~\ref{fig:age} (right). The Fornax sample is in between Sculptor and the Milky Way samples in metallicity; $\langle\text{[Fe/H]}_\textsl{Fnx}\rangle=-0.9\pm0.3$ \citep{Letarte10,Letarte18,Andrievsky17}. In spite of that, its [Y/Mg] values are higher than observed for $MW_\textsl{low}$ and most compatible with the highest metallicity bin, $MW_\textsl{high}$ ($\text{[Fe/H]>0}$). Furthermore, [Ba/Mg] is higher than all Milky Way data. We note that the Fornax data alone have been corrected for NLTE effects in Ba \citep{Andrievsky17}, but at the relevant stellar parameters, the corrections are small, $\approx\pm0.1$~dex \citep{Korotin15}. \citet{Andrievsky17} obtained lower Ba abundances than previously measured with the LTE approach \citep{Letarte10,Letarte18,Lemasle14}. Thus the inclusion of NLTE effects in Fornax brings [Ba/Mg] closer to the Milky Way values, but does not significantly affect the comparison. 

The obvious differences in Fig.~\ref{fig:age} between the Sculptor dSph, Fornax dSph, and the Milky Way make it clear that the correlation of [Y/Mg] and [Ba/Mg] with age is very dependent on the system in which the stars were formed. Furthermore, this dependence is not only on metallicity, but also on the overall SFH of the system. In addition, stars that have experienced very strong $r$-process events like CEMP-$r$ stars, or very strong $s$-process contribution like CEMP-$s$ stars (see Fig.~\ref{fig:ncapmg}) will naturally fall out of any average trend of [Y/Mg] or [Ba/Mg] with age. The use of these abundance ratios as chemical clocks is therefore complicated. As is clear from Fig.~\ref{fig:age}, any such abundance-age relation has to be calibrated in the context of the metallity and galaxy in question. However, if the relation of these ratios, [Y/Mg] and [Ba/Mg], with age are known within a given galaxy (or Galactic component), they can potentially be used to derive age estimates for stars (or groups of stars) with otherwise unknown ages.

                \begin{figure}
   \centering
   \includegraphics[width=\hsize-1cm]{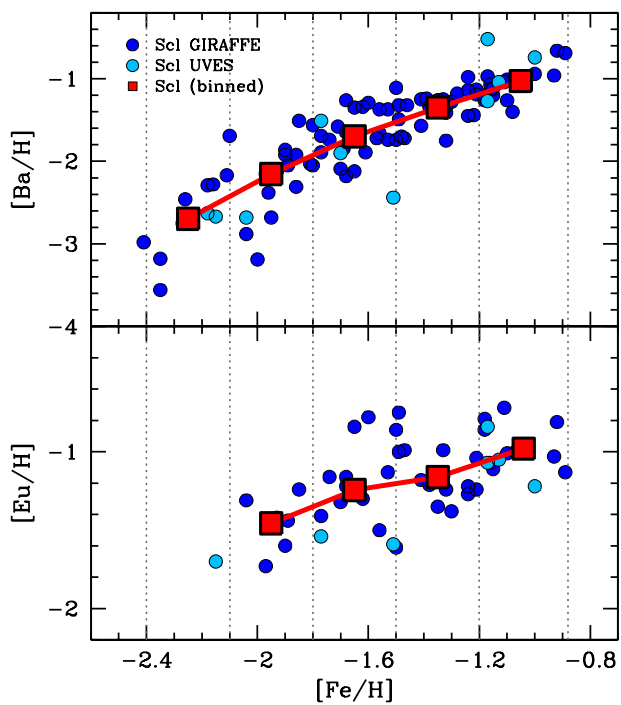}
      \caption{Build-up of [Ba/H] (top panel) and [Eu/H] (bottom panel) with [Fe/H]. Circles are individual Sculptor stars observed with GIRAFFE (blue) and UVES (light blue). Red squares are binned data, where the error of the mean is in all cases smaller than or comparable to the symbol size. Dotted lines show the edges of the [Fe/H] bins. The range of the y-axis is different in the two panels.
      }
         \label{fig:baheuh}
   \end{figure}   
   
                   \begin{figure}
   \centering
   \includegraphics[width=\hsize-1cm]{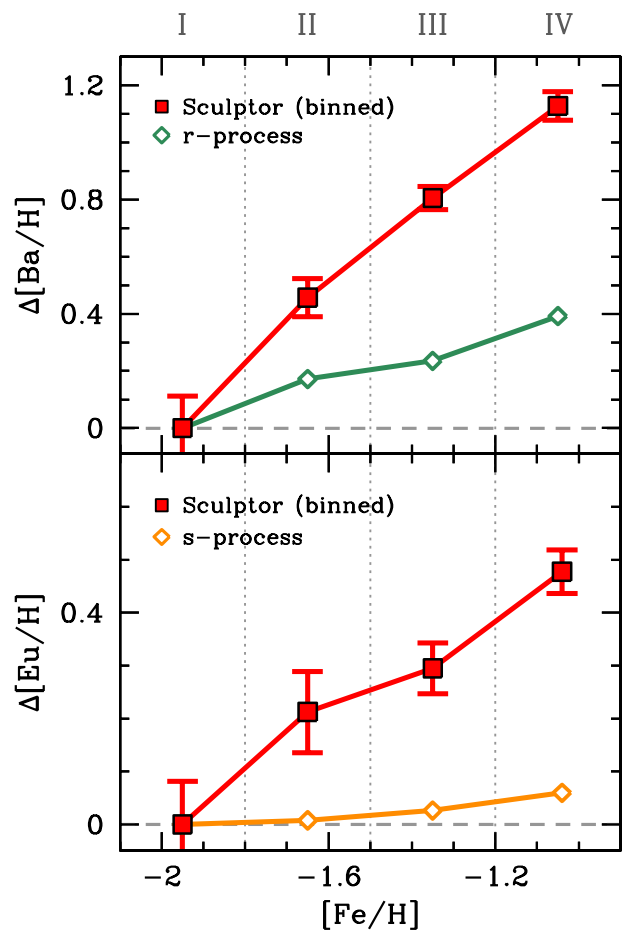}
      \caption{Increase in [Ba/H] (top panel) and [Eu/H] (bottom panel) with [Fe/H] (red squares) from the low-metallicity bin \textrm{I}~($-2.1<\text{[Fe/H]}\leq-1.8$) to more metal-rich bins: \textrm{II}~($-1.8<\text{[Fe/H]}\leq-1.5$), \textrm{III}~($-1.5<\text{[Fe/H]}\leq-1.2$), and \textrm{IV}~($-1.2<\text{[Fe/H]}\leq-0.9$), as indicated with dotted gray lines. Open diamonds are the contributions to [Ba/H] from the $r$-process (green), and to [Eu/H] from the $s$-process (orange). The range of the y-axis is different in the two panels.
      }
         \label{fig:dbaeu}
   \end{figure}   

Finally, we note that the absolute age scale of the SFH in Sculptor is uncertain, and is currently being debated. \citet{Weisz14} and \citet{Bettinelli19} derived a much shorter SFH in Sculptor ($\sim3$~Gyr), than was reported by \citet{Savino18}, and than we adopted here (Fig.~\ref{fig:sfhmdf}; \citealt{deBoer12}). Because studies other than that of \citet{deBoer12} do not provide ages for individual stars, it is difficult to adopt their age scale here in a robust quantitative way. However, the SFH of \citet{Weisz14} and \citet{Bettinelli19} would suggest that the age scale should be compressed and shifted toward older ages. This does not affect our conclusions, because [Y/Mg] with age in Sculptor would still be significantly different from that of the Milky Way, see Fig.~\ref{fig:age}. Furthermore, from Fig.~\ref{fig:ncapmg} it is clear that the [Y/Mg] of the Milky Way halo is on average lower than in Sculptor at the same metallicity, especially at $-2<\text{[Fe/H]}<-1.5$. In contrast to Sculptor, the Milky Way halo stars at this metallicity show no clear signs of SN type~Ia contribution, and are therefore probably older. This contradicts expectations, that is, increasing [Y/Mg] towards younger ages. Thus, chemical clocks are not universal, but are dependent on the SFH of each system.

\section{Build-up of the $s$- and $r$-process elements}\label{sec:buildup}

The differences in the abundance patterns of Ba and Eu in Fig.~\ref{fig:mgplot} show (as is commonly accepted) that these elements are predominately produced by different processes. The evolution of [Ba/H] and [Eu/H] with [Fe/H] in the Sculptor dSph is shown in Fig.~\ref{fig:baheuh}.  The abundances of Ba and Eu continue to increase throughout the chemical enrichment history of Sculptor. The absolute increase in [Ba/H] is, however, much more extreme than that of [Eu/H]. Thus the question becomes: \textit{How much of the increase in Eu is actually due to the $s$-process?} Although Eu is mostly produced in the $r$-process (94\% of the solar abundance according to \citealt{Bisterzo14}), the fraction depends on the SFH of the system, that is, the relative contributions of the $s$- and $r$-processes. Thus, a sufficiently strong $s$-process might affect the overall evolution of [Eu/H]. 

To ensure that the modest increase in [Eu/H] seen in Fig.~\ref{fig:baheuh} truly traces the $r$-process, we performed simple calculations of the relative contribution of the $r$-process to Ba and the $s$-process to Eu in Sculptor. The change in the observed abundances between two metallicity bins (shown in Fig.~\ref{fig:baheuh}) can be written as the combination of the contribution from the $r$- and $s$-processes into a mass of gas:

\begin{eqnarray}
\indent \Delta N_\text{Eu}^{obs}&=&\Delta N_\text{Eu}^{s}+\Delta N_\text{Eu}^{r}\label{eq:Eu}\\
\indent\Delta N_\text{Ba}^{obs}&=&\Delta N_\text{Ba}^{s}+\Delta N_\text{Ba}^{r}\label{eq:Ba}
\end{eqnarray}

\noindent Furthermore, we define
\begin{equation}
\indent\frac{\Delta N_\text{Ba}^{s}}{\Delta N_\text{Ba}^{r}}=\alpha, \hspace{0.5cm} \frac{\Delta N_\text{Eu}^{s}}{\Delta N_\text{Eu}^{r}}=\beta   \label{eq:alphabeta}
\end{equation}

\begin{equation}
\indent\frac{\beta}{\alpha}=\left(\frac{\Delta N_\text{Eu}^{s}}{\Delta N_\text{Ba}^{s}}\right)/\left(\frac{\Delta N_\text{Eu}^{r}}{\Delta N_\text{Ba}^{r}}\right)\label{eq:b/a}
\end{equation}
\noindent By combining Eq.~\ref{eq:Eu}, \ref{eq:Ba}, and \ref{eq:alphabeta} we obtain

\begin{equation}
\indent \frac{\Delta N_\text{Eu}^{obs}}{\Delta N_\text{Ba}^{obs}}=\frac{(1+\beta)}{(1+\alpha)}\left(\frac{\Delta N_\text{Eu}^{r}}{\Delta N_\text{Ba}^{r}}\right)\label{eq:obs}
\end{equation}

\noindent Solving for $\alpha$ in Eq.~\ref{eq:b/a} and \ref{eq:obs} gives
\begin{equation}
\indent \alpha=\frac{\Delta N_\text{Eu}^{r}/\Delta N_\text{Ba}^{r}-\Delta N_\text{Eu}^{obs}/\Delta N_\text{Ba}^{obs}}{\Delta N_\text{Eu}^{obs}/\Delta N_\text{Ba}^{obs}-\Delta N_\text{Eu}^{s}/\Delta N_\text{Ba}^{s}} \label{eq:final}
\end{equation}

\noindent This result is independent of the assumed mass of gas. Using the change in the observed average values in Sculptor from Fig.~\ref{fig:baheuh} and the ratios for the pure processes from \citet{Bisterzo14}, $\Delta N_\text{Eu}^{s}/\Delta N_\text{Ba}^{s}$ and $\Delta N_\text{Eu}^{r}/\Delta N_\text{Ba}^{r}$, we then calculated the relative contribution of the $s$- and $r$-process to both Ba and Eu. In Fig.~\ref{fig:dbaeu} we show the increase in [Ba/H] and [Eu/H] as a function of [Fe/H]. We adopted $\text{[Fe/H]}\approx-2$ as our initial conditions because this is the lowest metallicity bin from Fig.~\ref{fig:baheuh} with available Eu measurements. We also include in Fig.~\ref{fig:dbaeu} the increase in Ba as a result of the $r$-process, and of Eu as a result of the $s$-process, calculated using Eq.~\ref{eq:final}. 

As expected, the majority of Ba comes from the $s$-process, but with some contribution from the $r$-process ($\sim$13\% from the lowest to highest metallicity bin, I-IV). Even between the first two metallicity bins in Fig.~\ref{fig:dbaeu} (I-II), the Ba production is dominated by the $s$-process ($\sim$73\%).

 The total contribution of the $s$-process to [Eu/H] in Sculptor is $<0.1$~dex and corresponds to $\sim$8\% of the increase from the lowest (I) to the highest (IV) metallicity bins in Fig.~\ref{fig:dbaeu}. Thus, the $r$-process dominates the production of Eu at all times in the Sculptor dSph. Fig.~\ref{fig:dbaeu} clearly shows that the $r$-process was active throughout the final stages of the chemical evolution in the Sculptor dSph, even as star formation was dying out, see Fig.~\ref{fig:sfhmdf}.

                   \begin{figure*}
   \centering
   \includegraphics[width=\hsize-2cm]{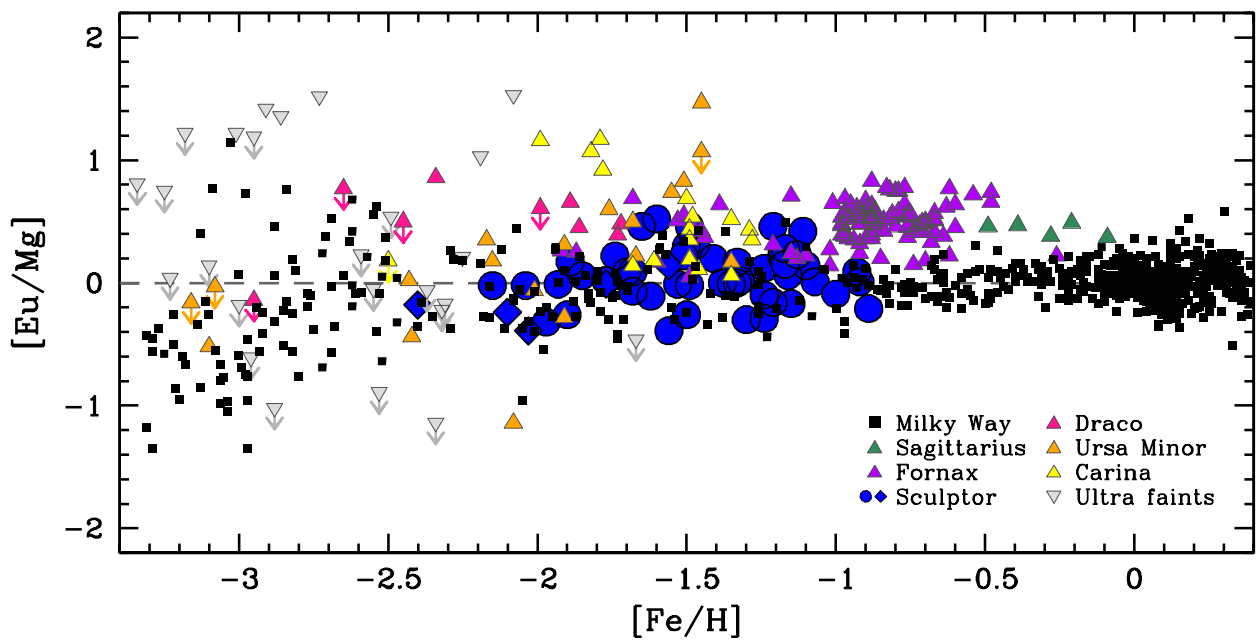}
      \caption{Measurements of [Eu/Mg] in individual stars in the Local Group. Confirmed CEMP-$s$ or CEMP-$s/r$ stars are not included. Black squares are Milky Way stars, and Sculptor is blue: circles are target stars (GIRAFFE and UVES), and diamonds are from the literature. Triangles are stars from other dwarf galaxies; included are (from highest to lowest stellar mass according to \citealt{McConnachie12}): \textit{Sagittarius (dark green):} \citealt{Bonifacio00,McWilliam13}; \textit{Fornax (violet):} \citealt{Shetrone03,Letarte10,Letarte18,Lemasle14}; \textit{Draco (pink):} \citealt{Shetrone01,Fulbright04,Cohen09};
       \textit{Ursa Minor (orange):} \citealt{Shetrone01,Sadakane04,Cohen10,Kirby12}; \textit{Carina (yellow):} \citealt{Susmitha17,Norris17}, which includes reanalysis of data presented in \citet{Shetrone03,Venn12,Lemasle12}; \textit{ultra-faint dwarf galaxies (gray):} \citealt{Frebel10UFD,Frebel14,RoedererKirby14,Ji16TucII,JiA16,JI19ApJ,HansenT17}. References for the Milky Way and Sculptor are listed in Fig.~\ref{fig:ncapmg}. 
      }
         \label{fig:eumg}
   \end{figure*}

\section{Origin of the $r$-process: Global view} \label{sec:global}

The observational data in the Milky Way and its satellite dwarf galaxies suggest that the $r$-process elements are produced in rare and extreme events, causing a large scatter of [Eu/Fe] ($\gtrsim2$~dex) at the lowest metallicities, $\text{[Fe/H]}\lesssim-2.5$ (e.g., \citealt{Francois07,Hansen12,Roederer14b,Roederer14,Ji16Nat,JiA16,JiA19,HansenT18,Sakari18}). In the following section we discuss what implications our new Sculptor results have on our understanding of the $r$-process when placed in context with literature data of individual stars in the Local Group.

\subsection{Eu ceiling in dwarf galaxies?}

\citet{Tsujimoto15b} suggested that there is an \textit{`Eu ceiling'} in dwarf spheroidal galaxies. Based on the available data at the time, they noted a plateau of $\text{[Eu/H]}\approx-1.3$ at $\text{[Fe/H]}>-2$ in three dSph galaxies: Draco, Carina, and Sculptor. The authors interpreted this as a sign of the rarity of $r$-process events in small systems, that is, after these galaxies enriched their gas up to $\text{[Fe/H]}\approx-2$ there was no more contribution from the $r$-process.

However, with our increased sample size of [Eu/H] measurements it is clear that this does not hold in the case of the Sculptor dSph, see Figs~\ref{fig:baheuh}, and \ref{fig:dbaeu}. The present data show that the mean [Eu/H] value increases steadily throughout the chemical evolution of Sculptor. The increase is modest, especially since at $\text{[Fe/H]}>-2$ the increase of [Fe/H] is mainly driven by SN type~Ia, which do not contribute to the $r$-process. Thus a significant number of stars, as presented here ($\approx$50), is needed to reveal the trend, especially considering the limited precision that is generally achieved in these faint distant systems. 

The dSph galaxies Draco and Carina have stellar masses $\sim3\cdot10^{5}$~M$_\odot$ \citep{McConnachie12}, which is about an order of magnitude lower than Sculptor. These galaxies therefore presumably also have fewer $r$-process events by an order of magnitude, likely $\lesssim$10, based on estimates of the $r$-process rate (e.g.,~\citealt{Ji16Nat}). It is therefore still possible that Draco and Carina exhibit such an Eu ceiling. However, this needs to be confirmed with a larger sample before strong conclusions are drawn. 

In addition to the Eu ceiling, \citet{Tsujimoto15b} suggested an \textit{`Eu jump'} in the Draco and Sculptor dSph, that is, a sudden increase in [Eu/H] at a fixed [Fe/H]. Their Fig.~1 shows a jump in [Eu/H] at low $\text{[Fe/H]}<-2.5$ in these galaxies. With the new increased sample size here, however, there is no evidence of this in the Eu abundances in Sculptor or in Ba (which at these metallicities is dominated by the $r$-process), see Fig.~\ref{fig:ncapmg}.\footnote{Fig.~1 in \citet{Tsujimoto15b} also has a small error where some upper limits in Sculptor are depicted as detections. This affects their conclusions.} In the case of Draco, the available data are still consistent with a possible Eu jump at $\text{[Fe/H]}\sim-3$, based on $\approx20$~stars with $\text{[Fe/H]}<-1.8$ \citep{Tsujimoto15a,Tsujimoto15b,Tsujimoto17}.

\subsection{Timescales of the $r$-process in the Local Group}

The available measurements of [Eu/Mg] in the Milky Way and its dwarf galaxy satellites are shown in Fig.~\ref{fig:eumg}. What is immediately obvious, is that the trend of [Eu/Mg] in the Milky Way is flat around the solar value, from the highest metallicities, $\text{[Fe/H]}\approx+0.5$ down to very low $\text{[Fe/H]}\approx-2.5$. The scatter in [Eu/Mg] increases towards the lowest metallicities, consistent with the idea that $r$-process events are rare and prolific, and sparsely sampled at the lowest [Fe/H] where only a handful of nucleosynthetic events might have contributed to the chemical abundance pattern of each star. 

The same trend is observed in the Sculptor dSph galaxy, with an average $\text{[Eu/Mg]}\approx0$ over the observed metallicity range $-2.5<\text{[Fe/H]}<-0.9$.  In general, nucleosynthetic events with significant time delay relative to ccSN have increasing [X/Mg], with time and [Fe/H] (e.g., Fig.~\ref{fig:mgplot}; \citealt{Skuladottir15b,Skuladottir17,Hill18}). Based on the data in the Milky Way and Sculptor shown in Fig.~\ref{fig:eumg}, there is thus no evidence that the $r$-process has a time delay relative to Mg, that is, \textit{the $r$-process enters the chemical evolution on the same timescales as the products of ccSN.}

Fig.~\ref{fig:eumg} shows quite extreme [Eu/Mg] ratios in the smallest ultra-faint dwarf galaxies (UFD) compared to those of the Milky Way or the larger dSph galaxies, that is, both higher and lower values. This is consistent with these small UFD either having experienced one $r$-process event or none (see also e.g., \citealt{Frebel16,Koch08,Koch13,JiA16}). 

The dSph galaxies, Draco, Ursa Minor, and Carina, all have a stellar mass on the order of a few times 10$^5$~M$_\odot$ \citep{McConnachie12}, that is, they are expected to have experienced only a handful of $r$-process events. Based on the available data, these three dSph galaxies have significantly different abundance distributions of [Eu/Mg], see Fig.~\ref{fig:eumg}. There is a very large scatter within Ursa Minor, where [Eu/Mg] increases toward higher metallicities, consistent with an $r$-process event that occurred late in its chemical evolution. The abundance trend in Draco can be explained with a similar $r$-process event. In contrast, Carina shows a decreasing trend with metallicity, indicating an $r$-process event when the ISM had reached $\text{[Fe/H]}\approx-2$. However, with the limited sample size it is unclear how uniform the [Eu/Mg] are at given [Fe/H] within each galaxy. Larger stellar samples are therefore crucial to fully understand the $r$-process enrichment in these small systems.

On the other hand, the trends of  [Eu/Mg] with [Fe/H] are more difficult to understand in the larger dSphs, Fornax and Sagittarius. Both galaxies have supersolar values of [Eu/Mg] that are not straightforward to explain, highlighting that our understanding of the $r$-process is by no means complete. At the lowest metallicities in Sagittarius, $\text{[Fe/H]}<-1$, the observed $\langle\text{[Eu/Ca]}\rangle=-0.04$ \citep{Hansen18} is consistent with the solar value, therefore it is likely that this enhancement of [Eu/Mg] occurred late in the history of the galaxy. However, we note that the abundance ratios of [Eu/Mg] in Fornax (which has more extended measurements) show no clear signs of increasing trends of [Eu/Mg] with [Fe/H].

At the higher metallicities in Sculptor, $\text{[Fe/H]}>-2$, the $r$-process seems to be well sampled, in the sense that the mean trend of [Eu/Mg] follows that of the Milky Way, and any signs of the stochasticity due to the rarity of the event are very subtle, if present. However, this is not the case for smaller dwarf galaxies, which show a very wide range of [Eu/Mg] values, see Fig.~\ref{fig:eumg}. We point out that in very small systems that experience less than 10 $r$-process events over their lifetime of several Gyr, a certain time delay is introduced, simply because the event is so rare. Therefore it is unlikely that these systems experience their first $r$-process event during the first $\sim$100 Myr of their star formation. This is important to take into account when time delays at the earliest times are interpreted. 

Finally, we note that our conclusions disagree with those of \citet{Duggan18}, who used the time delays of Ba, corrected for the $s$-process, to conclude that NSM were the dominant $r$-process source at early times in Sculptor, $\text{[Fe/H]}<-1.6$. The small number of stars in our dataset, combined with the large measurement errors on [Eu/Mg] do not allow us to draw robust conclusions about a possible slope in such a small metallicity range $-2.2\leq\text{[Fe/H]}\leq-1.6$. However, we point out that the $r$-process corrections applied in \citet{Duggan18} are uncertain for the following reasons: 1)~Because only a few stars in dwarf galaxies have [Ba/Eu] measurements, their corrections in Sculptor are only based on 7 stars below $\text{[Fe/H]}<-1.5$. 2)~These corrections are scaled on literature abundances, which are not homogeneously analyzed relative to each other, nor to the \citet{Duggan18} data. 3)~In Sculptor these corrections are based on data at $\text{[Fe/H]}>-2.2$. Still, corrected values of [Ba/Fe]$_r$ are shown down to $\text{[Fe/H]}=-2.8$. It is not clear how this extrapolation was made, nor how it affects the results. 4)~No error estimates for the corrections are provided, which makes it difficult to know at what level the final corrected abundances can be trusted. Given these uncertainties, we prefer drawing conclusions only from our own data, where no clear evidence of a time delay in the $r$-process is observed.

Furthermore, the interpretation that NSM are the dominant source of the $r$-process in dwarf galaxies only at early times has some inconsistencies. There is no obvious explanation for why NSM should only be important at early times, given that the one confirmed event, GW170807, was observed at present-day at low redshift. Also, when we consider that the [Eu/H] ratio in Sculptor continues to rise until the highest metallicities, $\text{[Fe/H]}\approx-1$, see Fig.~\ref{fig:baheuh}, it is evident that an additional process would be needed, and it is unclear what its role would be at early times. None of the known time delayed processes, such as SN type~Ia and AGB stars, contribute only at the earliest times and then stop, which means that such an argument would require very compelling evidence. In addition, there is no obvious explanation for why this early enrichment of NSM is only present in dwarf galaxies and not in the Milky Way, which has a flat trend of [Eu/Mg] at $-2.5<\text{[Fe/H]}<-1.5$. Finally this interpretation fails to explain why the [Eu/Mg] ratios in Sculptor are consistent with those of the Milky Way. This is not true for any ratios that trace different time delays relative to ccSN, such as [Ba/Mg] and [Fe/Mg]. 

The evolution of Ba with time of $\gtrsim10^8$~yr will always include some delayed contribution of the s-process, which is very challenging to correct for. To robustly remove the $s$-process contribution of Ba, a sizeable sample of [Eu/Ba] measurements needs to exist within each system, covering the same [Fe/H] as the Ba measurements. However, if such a sample exists, there is no need to use Ba, it can be done directly with Eu. \textit{We therefore do not recommend using time delays of Ba enrichment to conclude about time delays of the $r$-process.}

\subsection{The dominant $r$-process source in the Local Group}

The lack of evidence for a time delay of the $r$-process relative to ccSN raises doubts about the claim that NSM are the dominant (or only) nucleosynthetic source. The data are consistent with a very short (few million years) fixed time delay. However, NSM are expected to have a time-delay distribution on the form of $t^{-1}$ (e.g., \citealt{Cote19}, and references therein). In addition, around one-third of short gamma-ray bursts, which are believed to be associated with NSM, are found in early-type galaxies where star formation has ceased (\citealt{Berger14,Fong17}; and discussion in \citealt{Cote19}). Furthermore, the only identified host galaxy of an NSM is dominated by an old stellar population \citep{Blanchard17,Levan17,Pan17}.

One way for NSM to remain the dominant (only) source of $r$-process elements, is if there is some process that suppresses the expected increase in [Eu/Mg] based on their time-delay distributions. A few that have been proposed in the literature are listed here below.

\begin{itemize}
	\item \textit{Metallicity dependence:} \citet{Simonetti19} suggested that a metallicity-dependent probability of NSM, $\alpha_{NSM}$, could explain the observed trends of [Eu/Fe] with [Fe/H] in the Milky Way. That is, an increased probability to form NSM at low metallicities was able to alleviate the differences between models and data. However, with our new observations in Sculptor presented here, this hypothesis can be rejected. In the interval $-1<\text{[Fe/H]}<-2$, Sculptor and the Milky Way have the same $\text{[Eu/Mg]}\approx0$, even though the timescales of metallicity enrichment in Sculptor are much longer than in the Milky Way (see Fig.~\ref{fig:sfhmdf}).\\ 
	
    \item \textit{Two-phase ISM:} \citet{Schonrich19} proposed a two-phase ISM, where the products of ccSN are delayed from contributing to the chemical evolution. They assumed that $\text{[Eu/$\alpha$]}>0$ at $\text{[Fe/H]}>-1$ in the Milky Way cannot be explained by a one-phase ISM. We do not discuss the validity of this claim. However, Fig.~\ref{fig:eumg} clearly shows that [Eu/Mg] is always broadly consistent with zero. The interpretation of [Eu/Si] in the Milky Way disks is complicated by the fact that SN type~Ia are significant producers of Si but not of Eu (see, e.g., \citealt{Tsujimoto95,Iwamoto99}). Furthermore, the existing data require the effectiveness of this two-phase ISM to be metallicity independent, which seems unlikely.\\
    
    \item \textit{Natal kicks:} The explosion of the progenitor stars of the neutron star binary causes a so-called \textit{`natal velocity kick'}, which in some cases might cause the NSM to occur outside of the galaxy where the stars where formed (e.g.,~\citealt{Tauris17}). If this occurs preferentially to NSM with long timescales, it might explain the trend shown in Fig.~\ref{fig:eumg}. However, this would indicate that the Milky Way is as effective at keeping NS binaries in its potential well as Sculptor is, even though their total masses differ by several orders of magnitude, which is implausible. The theoretical expectation of natal kicks is even an argument against NSM as the dominant $r$-process source because this would lead to smaller [Eu/Mg] in dwarf galaxies than in the Milky Way, which is not observed, see Fig.~\ref{fig:eumg}. Recent simulations by \citet{Bonetti19} showed indeed that even small natal velocity kicks cause dwarf galaxies to lose a significant fraction of the NSM chemical products, while the effect is negligible for galaxies like the Milky Way.
    
\end{itemize}

At this point it is not possible to completely exclude NSM as an important $r$-process source, and that an unknown process counteracts the effect of the time-delay distribution. However, this hypothetical process has to be metallicity independent or have a dependence that goes hand in hand with the metallicity-dependence of ccSN. In addition, this unknown process has to act exactly the same in two very different galaxies, the Milky Way and the Sculptor dSph. There is currently no proposed process that fulfils these requirements. A more straightforward interpretation of the data is simply to assume that the $r$-process is dominated by massive stars. However, the possible production site is still being debated (e.g.,~\citealt{Cameron03,Winteler12,Nishimura15,Nishimura17,Siegel19a,Siegel19b}).

\section{Conclusions}

Spectra obtained with the ESO VLT/FLAMES spectrograph (both GIRAFFE and UVES) were used to study the heavy elements in 98 stars in the Sculptor dSph galaxy, covering the metallicity range $-2.4<\text{[Fe/H]}<-0.9$. We were able to measure the abundances of (number of stars): Mg (97) as an $\alpha$-element tracer of ccSN, and the heavy elements Y~(93), Ba~(98), La~(80), Nd~(81), and Eu~(51). 

By detailed study of our sample, we reached several conclusions regarding the $s$-process and the chemical clocks arising from its time delay. The influence of AGB stars on the chemical enrichment of Sculptor becomes apparent at $\text{[Fe/H]}\approx-2$, which corresponds to $\sim11.5$ Gyr ago, based on stellar ages from \citet{deBoer12}. For the first time, we probed the use of [Y/Mg] and [Ba/Mg] as chemical clocks in a galaxy different from the Milky Way. These abundance ratios have a strong correlation with age in Sculptor (increasing with younger ages), making them useful as chemical clocks. However, this trend is significantly offset from those observed in the Milky Way and the Fornax dSph galaxy. Thus we conclude that \textit{chemical clocks based on the delayed timescale of the $s$-process depend on both metallicity and environment.}

Our analysis of the heavy element abundances in the Sculptor dSph galaxy shows that the $r$-process contributes throughout the entire chemical evolution of Sculptor. That is, [Eu/H] continuously increases toward the highest [Fe/H], and this increase is dominated by the $r$-process ($>90\%$). The abundance ratios of [Eu/Mg] are flat around the solar value both in Sculptor and the Milky Way over a wide range of $\text{[Fe/H]}\gtrsim-2.5$. Because Mg is almost entirely created by ccSN ($\approx99\%$ of the solar abundance; \citealt{Tsujimoto95}), we can conclude that \textit{the $r$-process enters the chemical evolution of Sculptor on the same timescale as the products of massive stars.} Thus, there is no clear evidence that NSM are the dominant source of the $r$-process elements in dwarf galaxies at early or later times.

The only way that NSM can be the dominant (or only) nucleosynthetic site of the $r$-process is if a)~the predicted NSM time-delay distribution (based on theory and observations) is incorrect, and/or b)~there is some additional effect that causes the $r$-process to appear to be on the same timescale as massive stars. Any process that might hide the time-delay distribution of NSM needs to have exactly the same effect in the Milky Way as in Sculptor, a much smaller dwarf galaxy with a very different SFH. None of the suggested mechanisms (metallicity dependence, two-phased IMF, or natal kicks) can fulfil this condition. \textit{The consistent picture that is gained from [Eu/Mg] in the Sculptor dSph and the Milky Way strongly points to a $r$-process site associated with massive stars,} such as collapsars or magneto-rotationally driven SN. 

The abundances observed in smaller dwarf galaxies, Draco, Ursa Minor, Carina, and the UFD galaxies, are consistent with this interpretation. A more quantified study with detailed chemical evolution modeling is required to determine the upper limit of the NSM contribution, based on the available data. This is far beyond the scope of this paper. To avoid unnecessary complications, however, \textit{we strongly recommend that chemical evolution models focus on the [Eu/Mg] ratio}, which is independent of the influences of SN type~Ia, in contrast to other ratios, such as [Eu/Si], [Eu/Ca] and [Eu/Fe].

In contrast to the (relatively) clear picture that is obtained by focusing on the Milky Way and its smaller dwarf galaxy satellites, the largest dSphs, Sagittarius and Fornax, have supersolar $\text{[Eu/Mg]}\approx+0.5$ at $\text{[Fe/H]}\gtrsim-1$, which is difficult to explain. When we compare this with Sculptor it is tempting to explain the discrepancies with the longer SFH of the larger dwarf galaxies, that is, an additional source, that did not contribute to the chemical enrichment of Sculptor due to metallicity, or because Sculptor has not formed stars for the past $\sim$~6 Gyr. However, then it becomes challenging to explain why the Milky Way does not show evidence of the same unidentified source. Larger homogeneously analyzed samples in all dwarf galaxies, covering wide [Fe/H] ranges, are needed to fully understand the origin of the $r$-process. At the moment, however, this discrepancy remains a mystery.

\begin{acknowledgements}
 \'A.S.~acknowledges funds from the Alexander von Humboldt Foundation in the framework of the Sofja Kovalevskaja Award endowed by the Federal Ministry of Education and Research. A. C.~acknowledges funding from the Swiss National Science Foundation under grant P2GEP2\_184492. \'A.S.~thanks P.-C. Bonifacio, B. C{\^o}t{\'e}, and T.~Hansen, for useful advice and insightful suggestions. We also thank E.~Tolstoy and K.~Lind for careful reading of the manuscript. 
\end{acknowledgements}

\bibliography{heimildir}
\clearpage
\pagebreak

\begin{appendix}

\section{Comparison with previous abundance measurements}

\begin{figure}
   \centering
   \includegraphics[width=\hsize]{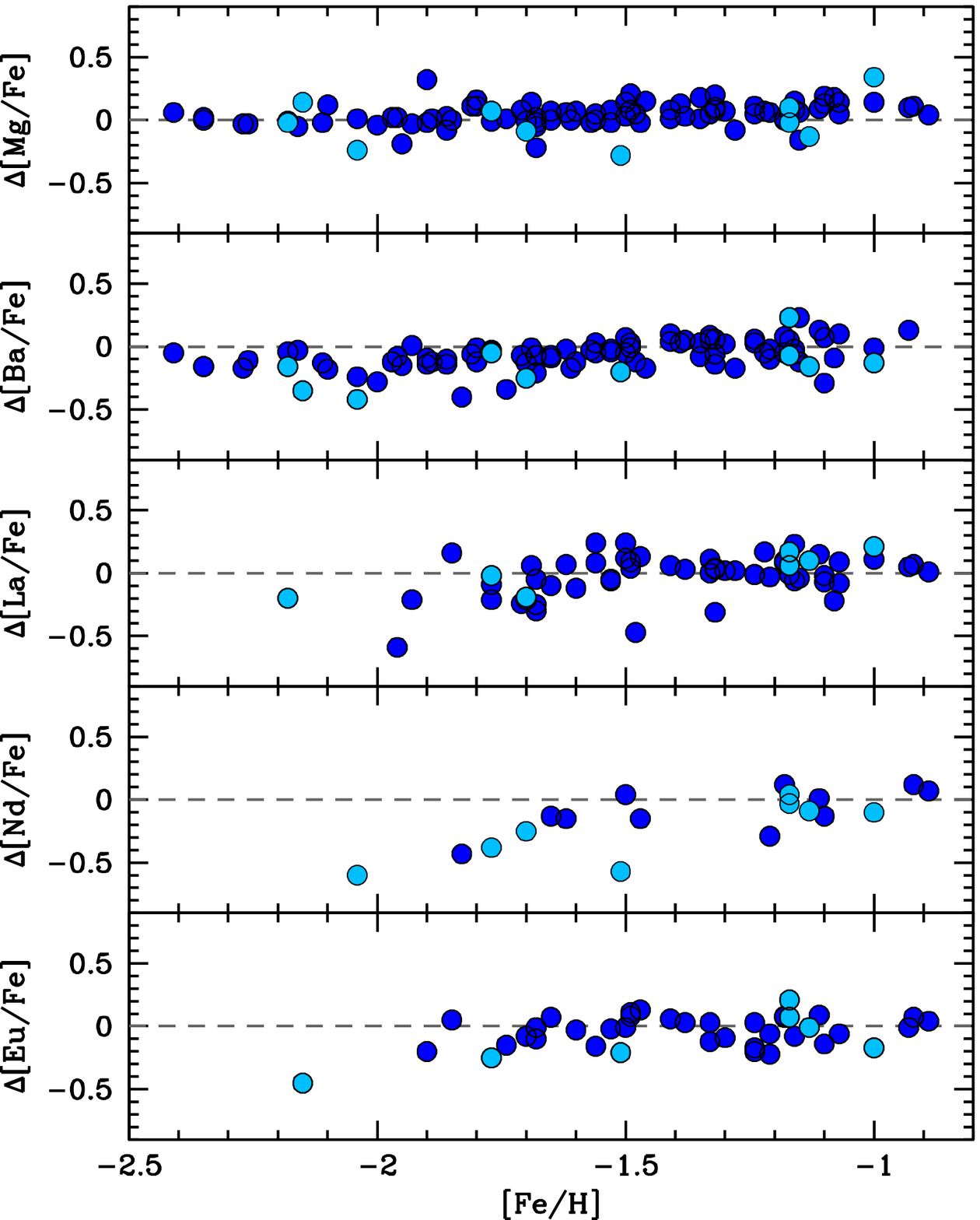}
      \caption{Comparison of measurements in this work and \citet{Hill18}. $\Delta\text{[X/Fe]}=\text{[X/Fe]}-\text{[X/Fe]}_\textsl{Hill}$.
      }
         \label{fig:van}
   \end{figure}

Fig.~\ref{fig:van} shows the comparison of the abundance analysis here with that of \citet{Hill18}. Overall, the abundance analysis is in good agreement, given the different methods. A clear exception are [La/Fe] and [Nd/Fe], where \citet{Hill18} used only weak lines in the redder settings, and therefore in some cases overestimated the abundances. By including the HR7 region, the abundance determination of these elements becomes much more robust.

\section{Online tables}

\setcounter{table}{0}
\begin{sidewaystable*}
\caption{\label{tab:abundances}The stellar parameters and measured abundances for our sample. The adopted solar abundances are: A(Fe)$_\odot=7.50$, A(Mg)$_\odot=7.58$, and A(Y)$_\odot=2.24$, A(Ba)$_\odot=2.13$, A(La)$_\odot=1.17$, A(Nd)$_\odot=1.50$, and A(Eu)$_\odot=0.51$ \citep{GrevesseSauval98}.
}
\centering
\scriptsize
\tabcolsep=0.15cm
\renewcommand{\arraystretch}{1.0}
\begin{tabular}{lccccccrccrccrccrccrccrc}
\hline\hline
Star	&	$T_\textsl{eff}  [K]$	&	$\log{g}$	&	$v_t$	&$	\text{[Fe/H]}	$&	$\delta_\text{[Fe/H]}$	&	$N_\text{Mg}$	&$	\text{[Mg/Fe]}	$&	$\delta_\text{[Mg/Fe]}$	&	$N_\text{Y}$	&$	\text{[Y/Fe]}	$&	$\delta_\text{[Y/Fe]}$	&	$N_\text{Ba}$	&$	\text{[Ba/Fe]}	$&	$\delta_\text{[Ba/Fe]}$	&	$N_\text{La}$	&$	\text{[La/Fe]}	$&	$\delta_\text{[La/Fe]}$	&	$N_\text{Nd}$	&$	\text{[Nd/Fe]}	$&	$\delta_\text{[Nd/Fe]}$	&	$N_\text{Eu}$	&$	\text{[Eu/Fe]}	$&	$\delta_\text{[Eu/Fe]}$	\\
\hline
ET0009$^{a}$	&	4171	&	0.2	&	2.2	&$	-1.68	$&	0.16	&	1	&$	0.52	$&	0.16	&	0	&$	...	$&	...	&	2	&$	-0.29	$&	0.18	&	2	&$	-0.09	$&	0.36	&	4	&$	-0.03	$&	0.12	&	0	&$	...	$&	...	\\
ET0013$^{a}$	&	4286	&	0.2	&	1.4	&$	-1.68	$&	0.21	&	1	&$	0.31	$&	0.27	&	0	&$	...	$&	...	&	2	&$	-0.50	$&	0.36	&	0	&$	...	$&	...	&	0	&$	...	$&	...	&	0	&$	...	$&	...	\\
ET0024	&	3897	&	0.0	&	2.2	&$	-1.24	$&	0.10	&	1	&$	0.11	$&	0.13	&	3	&$	-0.70	$&	0.16	&	2	&$	-0.21	$&	0.17	&	5	&$	-0.31	$&	0.05	&	18	&$	-0.28	$&	0.06	&	1	&$	-0.14	$&	0.21	\\
ET0026	&	4245	&	0.5	&	1.7	&$	-1.80	$&	0.16	&	1	&$	0.55	$&	0.17	&	3	&$	-0.50	$&	0.15	&	2	&$	-0.25	$&	0.21	&	3	&$	-0.10	$&	0.22	&	8	&$	-0.10	$&	0.08	&	0	&$	...	$&	...	\\
ET0027	&	4178	&	0.3	&	2.2	&$	-1.50	$&	0.13	&	1	&$	0.15	$&	0.17	&	3	&$	-0.72	$&	0.15	&	2	&$	-0.24	$&	0.16	&	4	&$	-0.09	$&	0.06	&	13	&$	-0.27	$&	0.07	&	1	&$	-0.04	$&	0.31	\\
ET0028	&	4085	&	0.3	&	2.0	&$	-1.22	$&	0.11	&	1	&$	-0.02	$&	0.22	&	3	&$	-0.72	$&	0.19	&	2	&$	-0.22	$&	0.22	&	7	&$	-0.12	$&	0.09	&	15	&$	-0.24	$&	0.05	&	0	&$	...	$&	...	\\
ET0031	&	4329	&	0.5	&	2.1	&$	-1.68	$&	0.17	&	1	&$	0.53	$&	0.19	&	3	&$	-0.39	$&	0.17	&	2	&$	0.42	$&	0.21	&	6	&$	-0.07	$&	0.06	&	12	&$	0.03	$&	0.07	&	1	&$	0.47	$&	0.25	\\
ET0033	&	4302	&	0.6	&	1.7	&$	-1.77	$&	0.16	&	1	&$	0.34	$&	0.23	&	3	&$	-0.43	$&	0.16	&	2	&$	0.08	$&	0.24	&	4	&$	0.02	$&	0.12	&	8	&$	0.14	$&	0.10	&	1	&$	0.38	$&	0.37	\\
ET0035$^{a}$	&	4390	&	0.0	&	1.5	&$	-1.95	$&	0.23	&	1	&$	0.58	$&	0.25	&	0	&$	...	$&	...	&	2	&$	-0.73	$&	0.36	&	0	&$	...	$&	...	&	0	&$	...	$&	...	&	0	&$	...	$&	...	\\
ET0039$^{a}$	&	4463	&	0.5	&	1.3	&$	-2.10	$&	0.25	&	1	&$	0.64	$&	0.24	&	0	&$	...	$&	...	&	2	&$	0.41	$&	0.24	&	0	&$	...	$&	...	&	0	&$	...	$&	...	&	0	&$	...	$&	...	\\
ET0043	&	4276	&	0.6	&	1.7	&$	-1.24	$&	0.16	&	1	&$	-0.15	$&	0.26	&	3	&$	-0.80	$&	0.21	&	2	&$	0.10	$&	0.21	&	3	&$	-0.18	$&	0.26	&	8	&$	-0.25	$&	0.08	&	1	&$	0.01	$&	0.41	\\
ET0048	&	4490	&	0.5	&	1.7	&$	-1.90	$&	0.19	&	1	&$	0.56	$&	0.20	&	3	&$	-0.55	$&	0.11	&	2	&$	0.04	$&	0.18	&	2	&$	-0.06	$&	0.16	&	4	&$	-0.18	$&	0.20	&	1	&$	0.32	$&	0.49	\\
ET0051	&	3971	&	0.5	&	1.7	&$	-0.92	$&	0.12	&	1	&$	0.10	$&	0.16	&	3	&$	-0.23	$&	0.18	&	2	&$	0.26	$&	0.06	&	6	&$	-0.33	$&	0.09	&	13	&$	-0.27	$&	0.07	&	1	&$	0.15	$&	0.18	\\
ET0054	&	4309	&	0.6	&	1.8	&$	-1.81	$&	0.16	&	1	&$	0.46	$&	0.17	&	3	&$	-0.44	$&	0.13	&	2	&$	-0.22	$&	0.24	&	2	&$	-0.07	$&	0.17	&	7	&$	-0.23	$&	0.08	&	0	&$	...	$&	...	\\
ET0057	&	4174	&	0.6	&	1.9	&$	-1.33	$&	0.13	&	1	&$	0.18	$&	0.22	&	3	&$	-0.49	$&	0.14	&	2	&$	-0.05	$&	0.21	&	7	&$	-0.02	$&	0.05	&	16	&$	-0.01	$&	0.06	&	1	&$	0.22	$&	0.18	\\
ET0059	&	4252	&	0.2	&	2.0	&$	-1.53	$&	0.16	&	1	&$	0.04	$&	0.18	&	3	&$	-0.92	$&	0.12	&	2	&$	-0.21	$&	0.23	&	3	&$	-0.26	$&	0.27	&	9	&$	-0.23	$&	0.12	&	0	&$	...	$&	...	\\
ET0060	&	4242	&	0.2	&	1.7	&$	-1.56	$&	0.15	&	1	&$	0.45	$&	0.16	&	3	&$	-0.70	$&	0.15	&	2	&$	-0.10	$&	0.27	&	2	&$	-0.19	$&	0.26	&	7	&$	-0.21	$&	0.07	&	1	&$	0.08	$&	0.33	\\
ET0062	&	4340	&	0.3	&	1.7	&$	-2.27	$&	0.18	&	1	&$	0.64	$&	0.15	&	3	&$	-0.61	$&	0.39	&	2	&$	-0.48	$&	0.31	&	0	&$	...	$&	...	&	0	&$	...	$&	...	&	0	&$	...	$&	...	\\
ET0063	&	4136	&	0.3	&	1.7	&$	-1.18	$&	0.19	&	1	&$	0.14	$&	0.20	&	3	&$	-0.67	$&	0.24	&	2	&$	-0.04	$&	0.21	&	6	&$	-0.26	$&	0.10	&	17	&$	-0.15	$&	0.07	&	1	&$	0.43	$&	0.24	\\
ET0064	&	4216	&	0.5	&	1.9	&$	-1.38	$&	0.14	&	1	&$	0.17	$&	0.23	&	3	&$	-0.35	$&	0.26	&	2	&$	0.04	$&	0.20	&	6	&$	-0.11	$&	0.05	&	13	&$	-0.07	$&	0.07	&	1	&$	0.21	$&	0.24	\\
ET0066	&	4266	&	0.4	&	1.9	&$	-1.30	$&	0.14	&	1	&$	0.22	$&	0.17	&	3	&$	-0.58	$&	0.14	&	2	&$	0.02	$&	0.16	&	5	&$	-0.20	$&	0.05	&	11	&$	-0.14	$&	0.12	&	1	&$	-0.04	$&	0.33	\\
ET0067	&	4281	&	0.5	&	1.7	&$	-1.65	$&	0.16	&	1	&$	0.35	$&	0.22	&	3	&$	-0.09	$&	0.20	&	2	&$	0.30	$&	0.24	&	7	&$	0.17	$&	0.05	&	15	&$	0.25	$&	0.05	&	1	&$	0.87	$&	0.11	\\
ET0069	&	4443	&	0.7	&	1.7	&$	-2.11	$&	0.20	&	1	&$	0.55	$&	0.21	&	3	&$	-0.17	$&	0.18	&	2	&$	-0.06	$&	0.23	&	0	&$	...	$&	...	&	0	&$	...	$&	...	&	1	&$	0.67	$&	0.46	\\
ET0071	&	4243	&	0.5	&	1.7	&$	-1.35	$&	0.14	&	1	&$	0.01	$&	0.22	&	3	&$	-0.44	$&	0.24	&	2	&$	0.09	$&	0.18	&	5	&$	-0.06	$&	0.07	&	13	&$	-0.09	$&	0.07	&	1	&$	0.03	$&	0.42	\\
ET0073	&	4370	&	0.8	&	1.7	&$	-1.53	$&	0.17	&	1	&$	0.41	$&	0.25	&	3	&$	-0.06	$&	0.21	&	2	&$	0.16	$&	0.22	&	5	&$	0.06	$&	0.08	&	8	&$	0.02	$&	0.07	&	1	&$	0.44	$&	0.26	\\
ET0083	&	4359	&	0.4	&	1.9	&$	-1.97	$&	0.18	&	1	&$	0.56	$&	0.17	&	3	&$	-0.36	$&	0.18	&	2	&$	-0.18	$&	0.16	&	3	&$	-0.01	$&	0.15	&	7	&$	-0.07	$&	0.09	&	1	&$	0.27	$&	0.40	\\
ET0094	&	4129	&	0.0	&	2.2	&$	-1.86	$&	0.15	&	1	&$	0.56	$&	0.21	&	3	&$	-0.96	$&	0.20	&	2	&$	-0.45	$&	0.19	&	4	&$	0.07	$&	0.22	&	4	&$	-0.16	$&	0.17	&	0	&$	...	$&	...	\\
ET0095	&	4307	&	0.2	&	1.9	&$	-2.16	$&	0.20	&	1	&$	0.52	$&	0.21	&	3	&$	-0.12	$&	0.10	&	2	&$	-0.12	$&	0.17	&	3	&$	0.05	$&	0.28	&	4	&$	0.30	$&	0.13	&	0	&$	...	$&	...	\\
ET0103	&	4250	&	0.5	&	2.0	&$	-1.21	$&	0.13	&	1	&$	0.13	$&	0.17	&	3	&$	-0.43	$&	0.35	&	2	&$	0.02	$&	0.15	&	7	&$	-0.21	$&	0.08	&	9	&$	-0.24	$&	0.09	&	1	&$	-0.01	$&	0.29	\\
ET0104	&	4343	&	0.8	&	1.7	&$	-1.62	$&	0.17	&	1	&$	0.43	$&	0.23	&	3	&$	-0.35	$&	0.22	&	2	&$	0.29	$&	0.19	&	5	&$	0.09	$&	0.10	&	11	&$	0.29	$&	0.10	&	0	&$	...	$&	...	\\
ET0109	&	4003	&	0.0	&	2.6	&$	-1.85	$&	0.11	&	1	&$	0.55	$&	0.09	&	3	&$	-0.34	$&	0.15	&	2	&$	0.34	$&	0.08	&	6	&$	0.04	$&	0.08	&	16	&$	0.11	$&	0.04	&	1	&$	0.65	$&	0.12	\\
ET0113	&	4285	&	0.2	&	1.8	&$	-2.18	$&	0.19	&	1	&$	0.44	$&	0.18	&	3	&$	-0.41	$&	0.20	&	2	&$	-0.11	$&	0.13	&	3	&$	0.06	$&	0.37	&	7	&$	0.09	$&	0.11	&	0	&$	...	$&	...	\\
ET0121	&	4462	&	0.4	&	1.9	&$	-2.35	$&	0.20	&	1	&$	0.43	$&	0.14	&	3	&$	-0.68	$&	0.28	&	2	&$	-0.83	$&	0.14	&	0	&$	...	$&	...	&	0	&$	...	$&	...	&	0	&$	...	$&	...	\\
ET0126	&	4186	&	0.7	&	1.7	&$	-1.11	$&	0.16	&	1	&$	-0.03	$&	0.20	&	3	&$	-0.38	$&	0.18	&	2	&$	0.07	$&	0.21	&	6	&$	-0.24	$&	0.09	&	12	&$	-0.09	$&	0.08	&	1	&$	0.44	$&	0.19	\\
ET0132	&	4321	&	0.9	&	1.7	&$	-1.50	$&	0.15	&	1	&$	0.34	$&	0.19	&	3	&$	-0.18	$&	0.23	&	2	&$	0.39	$&	0.14	&	7	&$	0.22	$&	0.05	&	11	&$	0.24	$&	0.04	&	1	&$	0.67	$&	0.14	\\
ET0133	&	4201	&	0.7	&	1.7	&$	-1.07	$&	0.15	&	1	&$	0.06	$&	0.22	&	3	&$	-0.42	$&	0.20	&	2	&$	-0.01	$&	0.21	&	6	&$	-0.23	$&	0.12	&	12	&$	-0.21	$&	0.05	&	1	&$	0.12	$&	0.27	\\
ET0137	&	3858	&	0.2	&	1.8	&$	-0.89	$&	0.18	&	1	&$	-0.03	$&	0.18	&	3	&$	-0.69	$&	0.19	&	2	&$	0.20	$&	0.10	&	6	&$	-0.30	$&	0.08	&	15	&$	-0.34	$&	0.07	&	1	&$	-0.18	$&	0.27	\\
ET0138	&	4205	&	0.4	&	2.1	&$	-1.70	$&	0.15	&	1	&$	0.29	$&	0.17	&	3	&$	-0.58	$&	0.20	&	2	&$	-0.39	$&	0.19	&	3	&$	-0.09	$&	0.21	&	9	&$	0.03	$&	0.11	&	1	&$	0.42	$&	0.35	\\
ET0139	&	4058	&	0.2	&	2.3	&$	-1.41	$&	0.11	&	1	&$	0.03	$&	0.12	&	3	&$	-0.79	$&	0.10	&	2	&$	-0.16	$&	0.17	&	6	&$	-0.23	$&	0.05	&	11	&$	-0.21	$&	0.06	&	1	&$	0.28	$&	0.17	\\
ET0141	&	4188	&	0.3	&	1.9	&$	-1.68	$&	0.15	&	1	&$	0.43	$&	0.17	&	3	&$	-0.45	$&	0.13	&	2	&$	0.03	$&	0.21	&	4	&$	0.00	$&	0.05	&	8	&$	0.03	$&	0.05	&	1	&$	0.55	$&	0.21	\\
ET0147	&	4261	&	0.0	&	1.8	&$	-1.15	$&	0.16	&	1	&$	-0.15	$&	0.14	&	2	&$	-1.43	$&	0.40	&	2	&$	-0.05	$&	0.28	&	1	&$	-0.39	$&	0.31	&	9	&$	-0.24	$&	0.11	&	0	&$	...	$&	...	\\
ET0150	&	4108	&	0.7	&	1.7	&$	-0.93	$&	0.11	&	1	&$	-0.19	$&	0.21	&	3	&$	-0.24	$&	0.23	&	2	&$	-0.03	$&	0.18	&	5	&$	-0.29	$&	0.08	&	8	&$	-0.34	$&	0.13	&	1	&$	-0.05	$&	0.30	\\
ET0151	&	4281	&	0.6	&	1.7	&$	-1.77	$&	0.16	&	1	&$	0.33	$&	0.19	&	3	&$	-0.49	$&	0.19	&	2	&$	-0.12	$&	0.26	&	3	&$	-0.03	$&	0.38	&	6	&$	0.03	$&	0.16	&	0	&$	...	$&	...	\\
ET0158	&	4452	&	0.9	&	1.4	&$	-1.80	$&	0.21	&	1	&$	0.50	$&	0.24	&	3	&$	-0.27	$&	0.24	&	2	&$	0.24	$&	0.29	&	2	&$	0.05	$&	0.29	&	7	&$	0.25	$&	0.06	&	0	&$	...	$&	...	\\
ET0160	&	4262	&	0.7	&	1.7	&$	-1.16	$&	0.14	&	1	&$	0.03	$&	0.24	&	3	&$	-0.70	$&	0.22	&	2	&$	0.03	$&	0.24	&	4	&$	0.04	$&	0.09	&	5	&$	-0.09	$&	0.06	&	1	&$	0.14	$&	0.24	\\
ET0163	&	4471	&	0.6	&	1.7	&$	-1.86	$&	0.21	&	1	&$	0.39	$&	0.21	&	3	&$	-0.72	$&	0.22	&	2	&$	-0.06	$&	0.23	&	2	&$	0.09	$&	0.25	&	4	&$	-0.10	$&	0.23	&	0	&$	...	$&	...	\\
ET0164	&	4454	&	0.6	&	1.8	&$	-1.89	$&	0.22	&	1	&$	0.28	$&	0.27	&	3	&$	-0.41	$&	0.20	&	2	&$	-0.16	$&	0.30	&	1	&$	-0.15	$&	0.17	&	4	&$	0.33	$&	0.17	&	1	&$	0.50	$&	0.40	\\
ET0165	&	4242	&	0.9	&	1.7	&$	-1.10	$&	0.17	&	1	&$	0.21	$&	0.19	&	3	&$	-0.62	$&	0.23	&	2	&$	0.09	$&	0.18	&	4	&$	-0.25	$&	0.05	&	11	&$	-0.04	$&	0.11	&	1	&$	-0.01	$&	0.50	\\
ET0166	&	4353	&	0.8	&	1.7	&$	-1.49	$&	0.15	&	1	&$	0.51	$&	0.19	&	3	&$	-0.53	$&	0.24	&	2	&$	0.00	$&	0.22	&	5	&$	0.10	$&	0.05	&	11	&$	0.06	$&	0.06	&	1	&$	0.53	$&	0.46	\\
ET0168	&	4292	&	0.5	&	1.7	&$	-1.10	$&	0.17	&	1	&$	-0.05	$&	0.20	&	3	&$	-0.33	$&	0.35	&	2	&$	-0.16	$&	0.30	&	6	&$	-0.23	$&	0.13	&	9	&$	-0.27	$&	0.13	&	1	&$	0.08	$&	0.37	\\
ET0173	&	3938	&	0.0	&	2.4	&$	-1.47	$&	0.10	&	1	&$	0.17	$&	0.20	&	3	&$	-0.61	$&	0.31	&	2	&$	-0.25	$&	0.13	&	3	&$	-0.13	$&	0.20	&	11	&$	-0.22	$&	0.07	&	1	&$	0.54	$&	0.25	\\
ET0198	&	4415	&	0.8	&	1.7	&$	-1.16	$&	0.17	&	1	&$	0.17	$&	0.19	&	3	&$	-0.47	$&	0.22	&	1	&$	0.09	$&	0.28	&	4	&$	-0.28	$&	0.08	&	6	&$	-0.18	$&	0.10	&	0	&$	...	$&	...	\\

\hline\hline
\multicolumn{7}{l}{$^{(a)}$ No HR7A spectrum available.}\\
\end{tabular}
\end{sidewaystable*} 

\clearpage
\pagebreak

\setcounter{table}{0}
\begin{sidewaystable*}
\label{tab:abundances2}

\caption{Continued. 
}
\centering
\scriptsize
\tabcolsep=0.15cm
\renewcommand{\arraystretch}{1.0}
\begin{tabular}{lccccccrccrccrccrccrccrc}
\hline\hline
Star	&	$T_\textsl{eff}  [K]$	&	$\log{g}$	&	$v_t$	&$	\text{[Fe/H]}	$&	$\delta_\text{[Fe/H]}$	&	$N_\text{Mg}$	&$	\text{[Mg/Fe]}	$&	$\delta_\text{[Mg/Fe]}$	&	$N_\text{Y}$	&$	\text{[Y/Fe]}	$&	$\delta_\text{[Y/Fe]}$	&	$N_\text{Ba}$	&$	\text{[Ba/Fe]}	$&	$\delta_\text{[Ba/Fe]}$	&	$N_\text{La}$	&$	\text{[La/Fe]}	$&	$\delta_\text{[La/Fe]}$	&	$N_\text{Nd}$	&$	\text{[Nd/Fe]}	$&	$\delta_\text{[Nd/Fe]}$	&	$N_\text{Eu}$	&$	\text{[Eu/Fe]}	$&	$\delta_\text{[Eu/Fe]}$	\\
\hline
ET0200	&	4370	&	1.0	&	1.7	&$	-1.49	$&	0.19	&	1	&$	0.28	$&	0.26	&	3	&$	-0.13	$&	0.22	&	2	&$	0.17	$&	0.19	&	7	&$	0.31	$&	0.09	&	11	&$	0.33	$&	0.08	&	1	&$	0.83	$&	0.28	\\
ET0202	&	4386	&	0.6	&	1.7	&$	-1.32	$&	0.19	&	1	&$	0.29	$&	0.13	&	2	&$	-0.92	$&	0.28	&	1	&$	-0.43	$&	0.45	&	3	&$	-0.11	$&	0.38	&	6	&$	-0.19	$&	0.18	&	0	&$	...	$&	...	\\
ET0206	&	4396	&	0.7	&	1.7	&$	-1.33	$&	0.17	&	1	&$	0.17	$&	0.24	&	3	&$	-0.55	$&	0.31	&	2	&$	0.08	$&	0.19	&	8	&$	0.03	$&	0.05	&	11	&$	-0.02	$&	0.11	&	1	&$	0.38	$&	0.27	\\
ET0232	&	4589	&	1.1	&	1.4	&$	-1.00	$&	0.18	&	1	&$	0.04	$&	0.19	&	3	&$	-0.56	$&	0.23	&	2	&$	0.06	$&	0.22	&	2	&$	-0.12	$&	0.40	&	7	&$	-0.22	$&	0.13	&	0	&$	...	$&	...	\\
ET0236	&	4547	&	1.0	&	1.7	&$	-2.41	$&	0.21	&	1	&$	0.48	$&	0.17	&	2	&$	-0.18	$&	0.33	&	2	&$	-0.57	$&	0.14	&	0	&$	...	$&	...	&	0	&$	...	$&	...	&	0	&$	...	$&	...	\\
ET0237	&	4430	&	0.6	&	1.7	&$	-1.61	$&	0.18	&	1	&$	0.34	$&	0.26	&	3	&$	-0.49	$&	0.21	&	2	&$	-0.28	$&	0.25	&	1	&$	0.21	$&	0.20	&	6	&$	-0.19	$&	0.21	&	0	&$	...	$&	...	\\
ET0238	&	4387	&	0.5	&	1.7	&$	-1.57	$&	0.17	&	1	&$	0.36	$&	0.22	&	3	&$	-0.64	$&	0.22	&	2	&$	-0.15	$&	0.21	&	3	&$	0.05	$&	0.39	&	5	&$	-0.31	$&	0.10	&	0	&$	...	$&	...	\\
ET0239	&	4526	&	0.8	&	1.7	&$	-2.26	$&	0.21	&	1	&$	0.60	$&	0.18	&	3	&$	-0.50	$&	0.25	&	2	&$	-0.20	$&	0.19	&	0	&$	...	$&	...	&	3	&$	0.43	$&	0.16	&	0	&$	...	$&	...	\\
ET0240	&	4343	&	0.8	&	1.5	&$	-1.15	$&	0.17	&	1	&$	0.21	$&	0.20	&	3	&$	-0.36	$&	0.30	&	2	&$	0.03	$&	0.43	&	2	&$	-0.28	$&	0.40	&	10	&$	-0.12	$&	0.11	&	1	&$	0.07	$&	0.39	\\
ET0241	&	4434	&	1.0	&	1.7	&$	-1.41	$&	0.17	&	1	&$	0.21	$&	0.24	&	3	&$	-0.71	$&	0.23	&	2	&$	0.16	$&	0.21	&	4	&$	0.04	$&	0.20	&	8	&$	0.02	$&	0.13	&	0	&$	...	$&	...	\\
ET0242	&	4371	&	0.7	&	1.7	&$	-1.32	$&	0.17	&	1	&$	0.06	$&	0.22	&	3	&$	-0.74	$&	0.19	&	2	&$	-0.09	$&	0.25	&	5	&$	-0.13	$&	0.06	&	11	&$	-0.09	$&	0.11	&	1	&$	0.13	$&	0.37	\\
ET0244	&	4433	&	0.8	&	1.7	&$	-1.24	$&	0.17	&	1	&$	0.12	$&	0.16	&	3	&$	-0.51	$&	0.24	&	2	&$	0.26	$&	0.21	&	5	&$	-0.11	$&	0.07	&	10	&$	-0.18	$&	0.10	&	1	&$	0.07	$&	0.44	\\
ET0270	&	4366	&	0.8	&	1.7	&$	-1.56	$&	0.16	&	1	&$	0.40	$&	0.19	&	3	&$	-0.52	$&	0.22	&	2	&$	0.19	$&	0.18	&	3	&$	0.20	$&	0.28	&	5	&$	0.07	$&	0.18	&	0	&$	...	$&	...	\\
ET0275	&	4478	&	1.0	&	1.7	&$	-1.21	$&	0.16	&	1	&$	-0.29	$&	0.19	&	3	&$	-1.28	$&	0.43	&	2	&$	0.08	$&	0.22	&	3	&$	-0.17	$&	0.21	&	7	&$	-0.20	$&	0.18	&	1	&$	0.17	$&	0.36	\\
ET0299	&	4704	&	1.3	&	1.7	&$	-1.83	$&	0.18	&	0	&$	...	$&	...	&	3	&$	-0.51	$&	0.28	&	2	&$	0.04	$&	0.15	&	2	&$	0.25	$&	0.25	&	3	&$	0.27	$&	0.19	&	0	&$	...	$&	...	\\
ET0300	&	4440	&	1.1	&	1.7	&$	-1.39	$&	0.20	&	1	&$	0.07	$&	0.24	&	2	&$	-0.66	$&	0.39	&	2	&$	0.15	$&	0.33	&	3	&$	0.20	$&	0.40	&	5	&$	0.14	$&	0.12	&	0	&$	...	$&	...	\\
ET0317	&	4434	&	0.8	&	1.7	&$	-1.69	$&	0.19	&	1	&$	0.34	$&	0.15	&	3	&$	-0.57	$&	0.19	&	2	&$	-0.16	$&	0.19	&	3	&$	0.29	$&	0.30	&	5	&$	-0.02	$&	0.13	&	0	&$	...	$&	...	\\
ET0320	&	4515	&	0.9	&	1.7	&$	-1.71	$&	0.21	&	1	&$	0.30	$&	0.16	&	3	&$	-0.55	$&	0.24	&	2	&$	0.13	$&	0.25	&	3	&$	0.02	$&	0.29	&	10	&$	0.17	$&	0.11	&	0	&$	...	$&	...	\\
ET0321	&	4360	&	0.3	&	1.7	&$	-1.93	$&	0.18	&	1	&$	0.52	$&	0.22	&	0	&$	-0.67	$&	0.24	&	2	&$	-0.19	$&	0.33	&	2	&$	-0.08	$&	0.28	&	3	&$	0.08	$&	0.36	&	1	&$	0.55	$&	0.39	\\
ET0322	&	4514	&	0.5	&	1.5	&$	-2.04	$&	0.27	&	1	&$	0.75	$&	0.15	&	3	&$	-0.55	$&	0.24	&	2	&$	-0.84	$&	0.28	&	0	&$	...	$&	...	&	0	&$	...	$&	...	&	1	&$	0.76	$&	0.31	\\
ET0327	&	4349	&	0.8	&	1.7	&$	-1.32	$&	0.16	&	1	&$	-0.02	$&	0.24	&	3	&$	-0.52	$&	0.36	&	2	&$	-0.01	$&	0.21	&	5	&$	-0.02	$&	0.14	&	8	&$	-0.02	$&	0.07	&	0	&$	...	$&	...	\\
ET0330	&	4476	&	0.7	&	1.5	&$	-2.00	$&	0.24	&	1	&$	0.36	$&	0.25	&	2	&$	-0.72	$&	0.37	&	2	&$	-1.19	$&	0.22	&	0	&$	...	$&	...	&	0	&$	...	$&	...	&	0	&$	...	$&	...	\\
ET0339	&	4340	&	0.8	&	1.7	&$	-1.08	$&	0.14	&	1	&$	-0.36	$&	0.21	&	3	&$	-0.87	$&	0.28	&	2	&$	-0.32	$&	0.25	&	3	&$	-0.27	$&	0.40	&	9	&$	-0.18	$&	0.11	&	0	&$	...	$&	...	\\
ET0342$^{b}$	&	4524	&	1.3	&	1.6	&$	-1.35	$&	0.20	&	1	&$	0.06	$&	0.27	&	0	&$	...	$&	...	&	2	&$	0.09	$&	0.32	&	2	&$	0.09	$&	0.35	&	0	&$	...	$&	...	&	0	&$	...	$&	...	\\
ET0350	&	4686	&	1.3	&	1.7	&$	-1.90	$&	0.21	&	1	&$	0.58	$&	0.17	&	2	&$	-0.51	$&	0.20	&	2	&$	-0.02	$&	0.24	&	1	&$	0.10	$&	0.14	&	1	&$	0.18	$&	0.40	&	0	&$	...	$&	...	\\
ET0354	&	4607	&	1.2	&	1.4	&$	-1.07	$&	0.20	&	1	&$	0.03	$&	0.23	&	2	&$	-0.82	$&	0.33	&	2	&$	0.09	$&	0.24	&	1	&$	-0.05	$&	0.37	&	4	&$	-0.19	$&	0.33	&	0	&$	...	$&	...	\\
ET0363	&	4552	&	1.1	&	1.4	&$	-1.28	$&	0.17	&	1	&$	0.08	$&	0.31	&	3	&$	-0.59	$&	0.40	&	2	&$	0.10	$&	0.35	&	5	&$	0.16	$&	0.13	&	4	&$	0.07	$&	0.18	&	0	&$	...	$&	...	\\
ET0369	&	4481	&	0.4	&	1.7	&$	-2.35	$&	0.20	&	1	&$	0.22	$&	0.19	&	3	&$	-0.60	$&	0.19	&	2	&$	-1.21	$&	0.15	&	0	&$	...	$&	...	&	0	&$	...	$&	...	&	0	&$	...	$&	...	\\
ET0373	&	4532	&	0.9	&	1.7	&$	-1.96	$&	0.21	&	1	&$	0.36	$&	0.21	&	3	&$	-0.82	$&	0.31	&	2	&$	-0.42	$&	0.24	&	2	&$	-0.06	$&	0.40	&	0	&$	...	$&	...	&	0	&$	...	$&	...	\\
ET0376	&	4320	&	1.0	&	1.7	&$	-1.17	$&	0.17	&	1	&$	-0.17	$&	0.29	&	3	&$	-0.61	$&	0.25	&	1	&$	0.20	$&	0.26	&	4	&$	-0.09	$&	0.10	&	5	&$	-0.35	$&	0.20	&	0	&$	...	$&	...	\\
ET0378	&	4308	&	0.7	&	1.6	&$	-1.18	$&	0.15	&	1	&$	0.11	$&	0.27	&	3	&$	-0.43	$&	0.24	&	2	&$	-0.08	$&	0.24	&	4	&$	-0.10	$&	0.21	&	10	&$	-0.16	$&	0.10	&	1	&$	0.39	$&	0.47	\\
ET0379	&	4486	&	0.8	&	1.7	&$	-1.65	$&	0.18	&	1	&$	-0.07	$&	0.28	&	3	&$	-1.14	$&	0.23	&	2	&$	-0.47	$&	0.26	&	2	&$	0.22	$&	0.25	&	0	&$	...	$&	...	&	0	&$	...	$&	...	\\
ET0382	&	4475	&	0.7	&	1.3	&$	-1.74	$&	0.23	&	1	&$	0.36	$&	0.26	&	3	&$	-0.42	$&	0.28	&	2	&$	0.00	$&	0.28	&	0	&$	...	$&	...	&	6	&$	0.14	$&	0.17	&	1	&$	0.66	$&	0.50	\\
ET0384	&	4497	&	1.1	&	1.4	&$	-1.46	$&	0.22	&	1	&$	0.07	$&	0.20	&	3	&$	-0.34	$&	0.31	&	2	&$	0.14	$&	0.31	&	4	&$	0.13	$&	0.23	&	4	&$	-0.02	$&	0.13	&	0	&$	...	$&	...	\\
ET0389	&	4394	&	0.8	&	1.5	&$	-1.60	$&	0.22	&	1	&$	0.30	$&	0.21	&	3	&$	-0.36	$&	0.32	&	2	&$	0.31	$&	0.22	&	4	&$	0.29	$&	0.08	&	9	&$	0.25	$&	0.14	&	1	&$	0.80	$&	0.50	\\
ET0392	&	4490	&	0.9	&	1.7	&$	-1.48	$&	0.20	&	1	&$	0.14	$&	0.36	&	2	&$	-0.84	$&	0.42	&	2	&$	-0.22	$&	0.32	&	3	&$	-0.08	$&	0.40	&	3	&$	0.09	$&	0.21	&	0	&$	...	$&	...	\\
\hline
\multicolumn{2}{l}{UVES fibres}			&		&		&$		$&		&		&$		$&		&		&$		$&		&		&$		$&		&		&$		$&		&		&$		$&		&		&$		$&		\\
\hline
UET0049	&	4255	&	0.2	&	2.3	&$	-2.18	$&	0.20	&	3	&$	0.39	$&	0.14	&	7	&$	-0.53	$&	0.06	&	3	&$	-0.45	$&	0.13	&	3	&$	0.00	$&	0.15	&	11	&$	-0.03	$&	0.09	&	0	&$	...	$&	...	\\
UET0065	&	4125	&	0.6	&	1.9	&$	-1.17	$&	0.14	&	3	&$	-0.02	$&	0.15	&	6	&$	-0.68	$&	0.07	&	3	&$	-0.10	$&	0.21	&	9	&$	-0.10	$&	0.05	&	23	&$	-0.01	$&	0.05	&	1	&$	0.10	$&	0.16	\\
UET0074	&	4340	&	0.9	&	1.2	&$	-1.17	$&	0.17	&	3	&$	0.08	$&	0.16	&	8	&$	-0.13	$&	0.06	&	3	&$	0.65	$&	0.23	&	9	&$	0.11	$&	0.05	&	17	&$	0.18	$&	0.06	&	1	&$	0.33	$&	0.21	\\
UET0082	&	4123	&	0.6	&	1.7	&$	-1.13	$&	0.15	&	3	&$	-0.13	$&	0.13	&	8	&$	-0.61	$&	0.05	&	3	&$	0.09	$&	0.22	&	11	&$	-0.17	$&	0.08	&	27	&$	-0.07	$&	0.05	&	1	&$	0.08	$&	0.15	\\
UET0112	&	4132	&	0.3	&	2.3	&$	-2.04	$&	0.17	&	3	&$	0.19	$&	0.22	&	7	&$	-0.47	$&	0.09	&	3	&$	-0.64	$&	0.22	&	6	&$	-0.01	$&	0.09	&	7	&$	-0.14	$&	0.10	&	0	&$	...	$&	...	\\
UET0127	&	4358	&	0.9	&	1.7	&$	-1.70	$&	0.17	&	3	&$	-0.02	$&	0.19	&	7	&$	-0.61	$&	0.08	&	3	&$	-0.20	$&	0.19	&	3	&$	-0.05	$&	0.25	&	10	&$	-0.11	$&	0.06	&	0	&$	...	$&	...	\\
UET0130	&	4426	&	0.6	&	1.7	&$	-2.15	$&	0.21	&	3	&$	0.50	$&	0.24	&	6	&$	-0.34	$&	0.11	&	3	&$	-0.52	$&	0.20	&	0	&$	...	$&	...	&	2	&$	0.10	$&	0.12	&	1	&$	0.45	$&	0.26	\\
UET0143	&	4281	&	0.4	&	1.5	&$	-1.77	$&	0.18	&	3	&$	0.24	$&	0.20	&	7	&$	-0.26	$&	0.08	&	3	&$	0.26	$&	0.27	&	6	&$	-0.03	$&	0.16	&	9	&$	0.09	$&	0.06	&	1	&$	0.23	$&	0.33	\\
UET0145	&	4286	&	0.5	&	1.8	&$	-1.51	$&	0.17	&	3	&$	-0.28	$&	0.21	&	5	&$	-1.00	$&	0.10	&	3	&$	-0.93	$&	0.18	&	0	&$	...	$&	...	&	5	&$	-0.71	$&	0.09	&	1	&$	-0.08	$&	0.42	\\
UET0152	&	4058	&	0.2	&	1.5	&$	-1.00	$&	0.16	&	3	&$	-0.11	$&	0.16	&	6	&$	-0.63	$&	0.07	&	3	&$	0.26	$&	0.29	&	9	&$	-0.29	$&	0.08	&	13	&$	-0.21	$&	0.10	&	1	&$	-0.22	$&	0.25	\\
\hline\hline
\multicolumn{10}{l}{$^{(b)}$ The HR7A spectrum had very low S/N, see \citet{Skuladottir17}.}\\
\end{tabular}
\end{sidewaystable*}

\begin{table}
\caption{\label{tab:linelist}The linelist for Mg, Y, Ba, La, Nd and Eu, including the central wavelength $\lambda$, excitation potential, $\chi_{ex}$, and the oscillator strength, $\log{gf}$. The last column is whether the line was used for the the GIRAFFE~(G) and/or the UVES~(U) spectra. 
}
\centering
\scriptsize
\renewcommand{\arraystretch}{0.9}
\begin{tabular}{l c c r c}
\hline\hline
El.	&	$\lambda$	&	$\chi_{ex}$	&	$\log{gf}$	&	G/U	\\
\hline								
\ion{Mg}{I}	&	5183.604	&	2.717	&$	-0.180	$&	U	\\
\ion{Mg}{I}	&	5528.405	&	4.346	&$	-0.620	$&	G/U	\\
\ion{Mg}{I}	&	5711.088	&	4.346	&$	-1.833	$&	U	\\
\hline									
\ion{Y}{I}	&	4854.252	&	1.887	&$	-0.030	$&	G/U	\\
\ion{Y}{II}	&	4854.861	&	0.992	&$	-0.380	$&	G/U	\\
\ion{Y}{II}	&	4883.682	&	1.084	&$	0.070	$&	G/U	\\
\ion{Y}{I}	&	4900.084	&	1.398	&$	-0.360	$&	G/U	\\
\ion{Y}{II}	&	4900.119	&	1.033	&$	-0.090	$&	G/U	\\
\ion{Y}{II}	&	5087.416	&	1.084	&$	-0.170	$&	U	\\
\ion{Y}{II}	&	5123.211	&	0.992	&$	-0.830	$&	U	\\
\ion{Y}{II}	&	5200.406	&	0.992	&$	-0.570	$&	U	\\
\ion{Y}{II}	&	5205.724	&	1.033	&$	-0.340	$&	U	\\
\ion{Y}{II}	&	5509.895	&	0.992	&$	-1.010	$&	U	\\
\hline									
\ion{Ba}{II}	&	5853.668	&	0.604	&$	-1.000	$&	U	\\
\ion{Ba}{II}	&	6141.713	&	0.704	&$	-0.076	$&	G/U	\\
\ion{Ba}{II}	&	6496.897	&	0.604	&$	-0.377	$&	G/U	\\
\hline									
\ion{La}{II}	&	4804.039	&	0.235	&$	-1.490	$&	G/U	\\
\ion{La}{II}	&	4808.996	&	0.235	&$	-1.400	$&	G/U	\\
\ion{La}{II}	&	4921.776	&	0.244	&$	-0.450	$&	G/U	\\
\ion{La}{II}	&	4986.819	&	0.173	&$	-1.300	$&	U	\\
\ion{La}{II}	&	5114.559	&	0.235	&$	-1.030	$&	U	\\
\ion{La}{II}	&	5122.988	&	0.321	&$	-0.850	$&	U	\\
\ion{La}{II}	&	5156.730	&	0.126	&$	-1.850	$&	U	\\
\ion{La}{II}	&	5290.818	&	0.000	&$	-1.650	$&	U	\\
\ion{La}{II}	&	5301.969	&	0.403	&$	-1.140	$&	U	\\
\ion{La}{II}	&	5303.528	&	0.321	&$	-1.350	$&	U	\\
\ion{La}{II}	&	5936.210	&	0.173	&$	-2.070	$&	U	\\
\ion{La}{II}	&	6262.287	&	0.403	&$	-1.220	$&	G/U	\\
\ion{La}{II}	&	6320.376	&	0.173	&$	-1.562	$&	G/U	\\
\ion{La}{II}	&	6390.477	&	0.321	&$	-1.410	$&	G/U	\\
\ion{La}{II}	&	6526.984	&	0.235	&$	-1.683	$&	G/U	\\
\ion{La}{II}	&	6774.268	&	0.126	&$	-1.708	$&	G/U	\\
\hline									
\ion{Nd}{II}	&	4783.828	&	0.064	&$	-1.553	$&	U	\\
\ion{Nd}{II}	&	4797.150	&	0.559	&$	-0.690	$&	G/U	\\
\ion{Nd}{II}	&	4799.420	&	0.000	&$	-1.450	$&	G/U	\\
\ion{Nd}{II}	&	4811.342	&	0.064	&$	-1.015	$&	G/U	\\
\ion{Nd}{II}	&	4825.478	&	0.182	&$	-0.420	$&	U	\\
\ion{Nd}{II}	&	4828.571	&	0.380	&$	-1.543	$&	G/U	\\
\ion{Nd}{II}	&	4859.026	&	0.321	&$	-0.440	$&	G/U	\\
\ion{Nd}{II}	&	4902.032	&	0.064	&$	-1.340	$&	G/U	\\
\ion{Nd}{II}	&	4914.380	&	0.380	&$	-0.700	$&	G/U	\\
\ion{Nd}{II}	&	4942.957	&	0.742	&$	-1.130	$&	G/U	\\
\ion{Nd}{II}	&	4943.899	&	0.205	&$	-1.514	$&	G/U	\\
\ion{Nd}{II}	&	4947.020	&	0.559	&$	-1.130	$&	G/U	\\
\ion{Nd}{II}	&	4949.011	&	0.631	&$	-1.483	$&	G/U	\\
\ion{Nd}{II}	&	4958.136	&	0.380	&$	-1.240	$&	G/U	\\
\ion{Nd}{II}	&	4959.115	&	0.064	&$	-0.800	$&	G/U	\\
\ion{Nd}{II}	&	4961.387	&	0.631	&$	-0.850	$&	G/U	\\
\ion{Nd}{II}	&	4987.160	&	0.742	&$	-0.790	$&	U	\\
\ion{Nd}{II}	&	4989.950	&	0.631	&$	-0.624	$&	U	\\
\ion{Nd}{II}	&	4998.541	&	0.471	&$	-1.166	$&	U	\\
\ion{Nd}{II}	&	5089.832	&	0.205	&$	-1.098	$&	U	\\
\ion{Nd}{II}	&	5092.788	&	0.380	&$	-0.610	$&	U	\\
\ion{Nd}{II}	&	5130.586	&	1.304	&$	0.450	$&	U	\\
\ion{Nd}{II}	&	5132.328	&	0.559	&$	-0.710	$&	U	\\
\ion{Nd}{II}	&	5212.360	&	0.205	&$	-0.960	$&	U	\\
\ion{Nd}{II}	&	5234.190	&	0.550	&$	-0.510	$&	U	\\
\ion{Nd}{II}	&	5249.576	&	0.976	&$	0.200	$&	U	\\
\ion{Nd}{II}	&	5276.869	&	0.859	&$	-0.668	$&	U	\\
\ion{Nd}{II}	&	5293.160	&	0.823	&$	0.100	$&	U	\\
\ion{Nd}{II}	&	5306.460	&	0.859	&$	-0.970	$&	U	\\
\ion{Nd}{II}	&	5311.450	&	0.986	&$	-0.420	$&	U	\\
\ion{Nd}{II}	&	5319.810	&	0.550	&$	-0.140	$&	U	\\
\ion{Nd}{II}	&	5361.165	&	0.559	&$	-1.480	$&	G/U	\\
\ion{Nd}{II}	&	5361.467	&	0.680	&$	-0.482	$&	G/U	\\
\ion{Nd}{II}	&	5385.888	&	0.742	&$	-0.860	$&	U	\\
\ion{Nd}{II}	&	5474.730	&	0.986	&$	-0.860	$&	U	\\
\ion{Nd}{II}	&	5485.696	&	1.264	&$	-0.120	$&	U	\\
\ion{Nd}{II}	&	6428.645	&	0.205	&$	-1.831	$&	G/U	\\
\ion{Nd}{II}	&	6514.959	&	0.182	&$	-1.883	$&	G/U	\\
\ion{Nd}{II}	&	6549.296	&	1.649	&$	-1.220	$&	U	\\
\ion{Nd}{II}	&	6549.525	&	0.064	&$	-2.010	$&	U	\\
\ion{Nd}{II}	&	6550.178	&	0.321	&$	-1.850	$&	U	\\
\ion{Nd}{II}	&	6740.078	&	0.064	&$	-1.526	$&	G/U	\\
\ion{Nd}{II}	&	6790.372	&	0.182	&$	-1.569	$&	G/U	\\
\hline									
\ion{Eu}{II}	&	6645.064	&	1.380	&$	0.120	$&	G/U	\\
\hline\hline								

\end{tabular}
\end{table}

\end{appendix}

\end{document}